\documentclass[pdflatex,sn-mathphys-num]{sn-jnl}
\usepackage{graphicx}%
\usepackage{multirow}%
\usepackage{amsmath,amssymb,amsfonts}%
\usepackage{amsthm}%
\usepackage{mathrsfs}%
\usepackage[title]{appendix}%
\usepackage{xcolor}%
\usepackage{textcomp}%
\usepackage{manyfoot}%
\usepackage{booktabs}%
\usepackage{algorithm}%
\usepackage{algorithmicx}%
\usepackage{algpseudocode}%
\usepackage{listings}%
\usepackage{nicefrac}
\usepackage{siunitx}
\usepackage{physics}
\usepackage{float}
\newcommand{\appropto}{\mathrel{\vcenter{
  \offinterlineskip\halign{\hfil$##$\cr
    \propto\cr\noalign{\kern2pt}\sim\cr\noalign{\kern-2pt}}}}}
\begin{document}

\title[Reflections on Future Problems in Cluster Science]
{Reflections on Future Problems in Cluster Science}

\author*[1]{\fnm{Klavs} \sur{Hansen}}\email{klavshansen@tju.edu.cn}
\affil*[1]{\orgdiv{Center for Joint Quantum Studies and Department of 
Physics, School of Science}, \orgname{Tianjin University}, \orgaddress{\street{Street}, 
\city{Tianjin}, \postcode{300072}, \country{China}}}

\author*[2]{\fnm{Vitaly} \sur{Kresin}}\email{Kresin@usc.edu}
\affil[2]{\orgdiv{Department of Physics and Astronomy}, \orgname{University of Southern California}, 
\orgaddress{\city{Los Angeles}, \postcode{90089-0484}, \state{California}, \country{USA}}}
\author[3]{\fnm{Ragheed} \sur{Alhyder}}\email{Ragheed.AlHyder@ist.ac.at}
\affil[3]{\orgname{Institute for Science and Technology Austria},
\orgaddress{\city{Klosterneuburg}, \postcode{3400}, \country{Austria}}}
\author[3]{\fnm{Mikhail} \sur{Lemeshko}}\email{mikhail.lemeshko@ist.ac.at}

\author[4]{\fnm{Michal} \sur{F\'{a}rn\'{i}k}}\email{michal.farnik@jh-inst.cas.cz}
\affil[4]{\orgdiv{J. Heyrovsk\'{y} Institute of Physical Chemistry}, 
\orgname{Czech Academy of Sciences}, \orgaddress{\street{Dolej\u{s}kova 2155/3}, \city{Prague}, \postcode{18223}, \country{Czech Republic}}}

\author[4]{\fnm{Juraj} \sur{Fedor}}\email{juraj.fedor@jh-inst.cas.cz}

\author[5,6]{\fnm{Piero} \sur{Ferrari}}\email{piero.ferrariramirez@ru.nl}
\affil[5]{\orgname{HFML-FELIX}\orgaddress{\street{Toernooiveld 7}, \city{Nijmegen}, 
\postcode{6525 ED}, \country{the Netherlands}}}
\affil[6]{\orgdiv{Institute for Molecules and Materials}, \orgname{Radboud University}
\orgaddress{\street{Heyendaalseweg 135}, \city{Nijmegen}, \postcode{6525 AJ}, 
\country{the Netherlands}}}
\author[5,6]{\fnm{Laura X.} \sur{Worutowicz}}
\author[5,6]{\fnm{Rick J.} \sur{Louwerse}}
\author[5,6]{\fnm{Denis} \sur{Kiawi}}
\author[5,7]{\fnm{Laurens B. F. M.} \sur{Waters}}
\affil[7]{\orgdiv{Department of Astrophysics, IMAPP}, \orgname{Radboud University}, 
\orgaddress{\city{Nijmegen}, \country{The Netherlands}}}
\author[8]{\fnm{Sandra M.} \sur{Lang}}
\affil[8]{\orgdiv{Institute of Surface Chemistry and Catalysis}, \orgname{University of Ulm}, 
\orgaddress{\postcode{89069}, \city{Ulm}, \country{Germany}}}
\author[5,6]{\fnm{Joost M.} \sur{Bakker}}
\author[9]{\fnm{Bernd} \spfx{von} \sur{Issendorff}}\email{bernd.von.issendorff@physik.uni-freiburg.de}
\affil[9]{\orgdiv{Physics Institute}, \orgname{University of Freiburg}, 
\orgaddress{\street{Hermann-Herder-Str. 3}, \city{Freiburg}, \postcode{79104}, \country{Germany}}}

\author[10]{\fnm{Wei} \sur{Kong}}\email{Wei.Kong@oregonstate.edu}
\affil[10]{\orgdiv{Department of Chemistry}, \orgname{Oregon State University},
\orgaddress{\city{Corvallis}, \postcode{97331}, \state{Oregon}, \country{USA}}}

\author[11]{\fnm{Jannik} \sur{Mehmel}}\email{jannik.mehmel@tu-darmstadt.de}
\author[11]{\fnm{Rolf} \sur{Sch\"afer}}
\affil[11]{\orgdiv{Eduard-Zintl-Institute}, \orgname{Technical University of Darmstadt}, 
\orgaddress{\street{Peter-Gr\"unberg-Stra{\ss}e 8}, \city{Darmstadt}, \postcode{64287}, 
\country{Germany}}}

\author[12,13]{\fnm{Sebastian} \sur{Pedalino}}\email{sebastian.pedalino@univie.ac.at}
\author[12,13]{\fnm{Bruno E.} \sur{Ram\'irez-Galindo}}
\author[12,13]{\fnm{Richard} \sur{Ferstl}}
\author[12,13]{\fnm{Severin} \sur{Sindelar}}
\author[12]{\fnm{Stefan} \sur{Gerlich}}
\author[12]{\fnm{Markus} \sur{Arndt}}\email{markus.arndt@univie.ac.at}
\affil[12]{\orgdiv{University of Vienna}, \orgname{Faculty of Physics}, 
\orgaddress{\street{Boltzmanngasse 5}, \city{Vienna}, \postcode{1090},  \country{Austria}}}
\affil[13]{\orgdiv{University of Vienna}, \orgname{Vienna Doctoral School of Physics}, 
\orgaddress{\street{Boltzmanngasse 5}, \city{Vienna}, \postcode{1090},  \country{Austria}}}

\author[14,15]{\fnm{Scott G.} \sur{Sayres}}\email{ssayres@asu.edu}
\affil[14]{\orgdiv{School of Molecular Sciences}, \orgname{Arizona State University},
\orgaddress{\city{Tempe}, \postcode{85287}, \state{Arizona}, \country{USA}}}
\affil[15]{\orgdiv{Biodesign Center for Applied Structural Discovery}, 
\orgname{Arizona State University},
\orgaddress{\city{Tempe}, \postcode{85287}, \state{Arizona}, \country{USA}}}

\author[16]{\fnm{Lai-Sheng} \sur{Wang}}\email{lai-sheng{\_}wang@brown.edu}
\affil[16]{\orgdiv{Department of Chemistry}, \orgname{Brown University}, 
\orgaddress{\city{Providence}, \postcode{02902}, \state{Rhode Island}, \country{USA}}}

\abstract{This article is a collection of contributions from speakers at the 
2025 DEAMN workshop at the Majorana Centre in Erice.
Not ordinary contributions to a conference proceeding, this gives a new and 
different perspective on the work done by the workshop participants.}


\maketitle

\section*{Introduction}

This article originated with a proposal to speakers at the May 2025 workshop on 
Dynamics of Electrons in Atomic and Molecular Nanoclusters.  
The workshop was hosted by the School of Solid State Physics (Prof. Giorgio Benedek, 
director) at the Ettore Majorana Centre for Scientific Culture in Erice (Sicily, 
Italy) and focused primarily on studies of free isolated nanoclusters. 
In the call we wrote:

\textit{Many of us may have had the experience of meeting visitors or other 
scientists, who listen to us describe our research and then ask: "So, what 
questions are the most exciting and promising in your field?"
Of course, the tempting response is "the ones on which I’m working at the moment." 
But actually it would be useful for ourselves, for graduate students and postdocs, 
and for outside colleagues, to have a sampling of your views about that question.}

\textit{Therefore what we'd like to propose is not a multi-authored "status of the field" 
review, but a collection of individual short reflections. 
We would like to invite each one of you to contribute a short note on what you 
view as really interesting, open, and fundamental problems or issues in the field 
that are worth pursuing. 
In other words, if you had plenty of freedom and resources, what ambitious 
problems would you love to explore in cluster science and its adjacent fields?}

The concept was welcomed by the editors of European Physical Journal D, and a number 
of authors have found the time to contribute to the article you see here. 
The wide range of topics represented shows the span of fields covered by the title 
of the workshop, and in a broader sense by the cluster community. 
We wish you a pleasant reading experience.

Klavs Hansen and Vitaly Kresin\\
Directors of the Workshop

\noindent
\vspace{1cm}
	
\section*{Molecules in Quantum Materials: What Questions Will Move the Field?}
\label{Lemeshko}
\noindent
Ragheed Alhyder and Mikhail Lemeshko

\subsection*{Introduction}
Contemporary quantum materials based on molecular platforms resist intuition because 
their governing physics spans many length and time scales~\cite{keimer2017}. 
Yet, many observed phenomena in these materials can be captured by minimal 
theoretical models that keep only a few essential degrees of freedom and symmetries, 
while remaining analytically transparent and experimentally testable.
This perspective underpins a range of concrete settings where our group has made 
recent progress, and where we see exciting opportunities for future work. 
A unifying theme is molecules, which unlike atoms, have an extended structure 
that supports rotational, vibrational, and spin–orbit dynamics. 
These additional degrees of freedom, while increasing complexity, give rise to 
rich many-body landscapes and enables coupling to external fields with nontrivial 
spatial or polarization structure. This results in a diverse network of interactions 
and a broader range of physical phenomena, making molecular systems fertile ground 
for theoretical exploration~\cite{Krems2018,koch2019,ruttley2025}.

Examples include hybrid organic–inorganic perovskites (HOIPs), where soft lattices 
and reorientable molecular units dictate charge transport, recombination, and 
polaron formation~\cite{brenner2016a,findik2021,biliroglu2025b}. Another is 
chirality-induced spin selectivity (CISS), where molecular handedness couples 
electron motion and spin in unexpectedly robust ways that call for minimal models 
beyond band-structure intuition~\cite{Evers2022}.
The same molecular degrees of freedom that shape transport and spin phenomena 
in condensed-phase systems can also be exploited to drive and control molecular 
dynamics. 
In particular, light with spatial or polarization structure can exchange angular 
momentum with molecular rotation, enhancing otherwise forbidden rovibrational 
transitions and coupling to center-of-mass motion~\cite{Maslov2024}. 
Building on that control, when molecules are driven by periodic laser 
pulses, their rotational spectra map to synthetic lattices with Dirac 
cones, enabling tunable topological charges in angular momentum 
space~\cite{Karle2023TopologicalKickedMolecules}. 
Microwave dressing of ultracold molecules further complements these schemes 
by reshaping internal structure and enabling long-range interactions to be 
captured in few-parameter effective Hamiltonians~\cite{Gorshkov2008,Fulin2023}.

\subsection*{Hybrid Perovskites and Rotational–Electronic Coupling} 
Hybrid organic–inorganic perovskites (HOIPs) offer a paradigmatic case for minimal 
modeling in molecular materials. Their soft, polar lattices harbor reorientable 
molecular units whose low-energy rotational modes couple strongly to charge 
carriers and lattice vibrations, enabling long carrier lifetimes, slow recombination, 
and large polaron formation~\cite{johnston2016,martiradonna2018,jena2019, wang2021}. 
Since these collective effects derive from a small set of microscopic ingredients, 
namely rotations, dipoles, and soft phonons, they are amenable to description by 
few-parameter Hamiltonians, providing a powerful framework for studying HOIPs. 
These models are used to infer scaling relations between reorientation dynamics, 
lattice softness, and macroscopic observables including mobility, recombination 
rates, and optical response. 
They also offer a natural way to analyze how local symmetry breaking, for example 
through external fields or spontaneous relaxation mechanisms reshape charge dynamics. 
\\Within this framework, charge–rotor coupling can induce ferroelectric ordering 
and modify carrier mobility and coherence~\cite{Koutentakis2023RotorLattice}, while 
in two dimensions it stabilizes domain-wall polarons with carriers localized along 
polarization boundaries~\cite{Kluibenschedl2025DomainWallPolarons}. 
In addition, spin–electric coupling provides an extra channel linking lattice 
dynamics to electronic spin structure~\cite{Volosniev2023SpinElectricPerovskites}, 
further enriching the effective low-energy description of these materials.

Looking forward, open questions concern how these mechanisms interplay with 
excitonic degrees of freedom, how higher-order electron–phonon couplings shape 
coherence and transport, and how minimal models can be systematically extended 
to capture these effects under realistic conditions. Developing such a framework 
could clarify the microscopic origin of several observations and guide the 
design of tailored perovskite architectures for optoelectronic and quantum 
applications.

\subsection*{Chirality and Spin}\label{II}

Chirality-induced spin selectivity (CISS) illustrates how molecular geometry alone 
can influence electron spin. Electrons traversing a chiral molecule can acquire spin 
polarization even without magnetic fields, an effect observed across diverse 
molecular systems and persisting at ambient temperatures~\cite{Evers2022}.

Current models stipulate that CISS emerges from coupling between molecular geometry 
and spin–orbit interactions. The degree of spin filtering depends on parameters 
such as spin-orbit coupling strength, molecular geometry, and contact coupling, 
yet the large polarizations detected in experiments remain an active area of 
research where many observations still need to be understood. 
Symmetry breaking from substrates, dissipation, or environmental interactions can 
further modulate the effect. Minimal models serve to clarify which microscopic 
ingredients are essential to explain the experimental observations in such complex 
systems. Examples include analytic treatments of chiral spin 
coupling~\cite{ghazaryan2020} and quantum transport under chiral molecular 
potentials~\cite{AlHyder2025Quantum}. 
To clearly distinguish genuine chirality-induced effects, related work has examined 
how achiral dipoles on ferromagnetic substrates can influence magnetization via 
Rashba-like coupling~\cite{AlHyder2023Achiral}.

Remaining challenges include determining how far minimal models can be extended to 
predict and control spin polarization in realistic chiral settings, and assessing 
their robustness when microscopic structure, disorder, or dissipation are 
explicitly included. 
Minimal models may guide predictions of chiral symmetry 
breaking in quantum platforms such as Josephson junctions embedding chiral molecules,
or describe how magnetic textures like skyrmions form or evolve on 
chiral-functionalized substrates. 
More broadly, the goal is to distill the 
essential CISS mechanisms into a compact parameter set valid across molecular 
platforms, and thereafter to optimize external perturbations, fields or geometry, 
for systematic enhancement, reversal, or switching of spin selectivity in a 
controlled, quantitative way.

\subsection*{Control of Molecular Rotations and Topology}

The ability to manipulate molecular rotation with tailored electromagnetic 
fields provides a versatile route to control quantum degrees of freedom with 
minimal ingredients. 
Structured light fields, carrying spatial or polarization structure, can 
exchange angular momentum with molecules through well-defined selection rules, 
enabling precise steering of rotational and rovibrational 
dynamics~\cite{Maslov2024}. 
At dipole order, polarization dictates rotational transfer, while at quadrupole 
and higher orders, spatial gradients enhance otherwise forbidden transitions, 
allowing efficient imprinting and amplification of rotational structure. 
This establishes a clean, few-parameter interface between light and molecular motion. 
More specifically, it enables us to address a plethora of questions pertaining 
to the usefulness of structured light in assessing the structure and steering the 
equilibrium and dynamical properties of molecular ensembles. 
For instance, an important paradigm connecting to subsection Chirality and Spin
is whether 
structured light can provide a sensitive probe of the enantiomeric purity of 
chiral molecules~\cite{hrast2025bottomupanalysisrovibrationalhelical}. 
Further, orbital angular momentum spectroscopy might enable a detailed tuning 
of the potential landscape and interactions of ensembles of ultracold 
atoms~\cite{Chen2018}, molecules and even to extensive objects such as 
optical-tweezers suspended nanoparticles~\cite{Gieseler2012}.

Light-molecule interactions can be further exploited through strong off-resonant 
light pulses that induce molecular alignment and orientation. 
In this regime, 
the strong field creates an effective dipole moment to which the field couples 
non-linearly~\cite{friedrich1995alignment,Karle2022ModelingKicks}. 
While this mechanism has numerous applications, one particularly intriguing 
approach involves periodic pulse sequences that generate a Dirac comb in both 
time and frequency domains~\cite{stummer2020programmable}. 
From a theoretical perspective, these phenomena can be understood through 
Floquet theory of angular momentum lattices~\cite{Karle2023TopologicalKickedMolecules}. 
Driving molecules in this manner produces topological Floquet bands that host 
symmetry-protected Dirac cones, whose topological charge and position can 
be tuned using simple experimental parameters such as pulse period, intensity, 
and polarization. 
These topological features manifest in observables like alignment traces and 
give rise to edge-like rotational states that can be created, moved, and 
annihilated through specific pulse sequences.

A key open direction is to extend these schemes beyond single-rotor physics: 
understanding their stability against interactions, disorder, and dissipation,
and developing minimal models for collective behavior in molecular arrays. 
Ultimately, structured fields may provide a route to engineer and stabilize 
topological states in driven molecular ensembles, bridging microscopic control 
and emergent many-body phenomena.

\subsection*{Field-Linked Molecules: Microwave Dressing and tailoring interactions} 

Microwave dressing of polar molecules near a field-linked resonance provides a 
powerful route to engineer tunable long-range interactions with minimal microscopic 
ingredients. 
By coupling internal rotational states through near-resonant fields, shallow 
field-linked dimers can be formed whose binding energy, interaction range, 
and angular structure are determined by only a few experimentally accessible 
parameters~\cite{Gorshkov2008,sundar2018a}. This tunability makes field-linked 
molecules a versatile platform for realizing strongly interacting dipolar 
systems with controllable symmetries.
Recent theoretical work has developed minimal models that capture the essential 
features of these resonances, enabling analytic control over interaction 
anisotropy, effective range, and energy dependence~\cite{Fulin2023,chen2023,li2025tunable}. 

Many questions remain in this nascent field, beginning with which minimal 
descriptions remain valid when field-linked states are embedded in complex 
many-body environments. 
A key goal is to translate two-body controllability into collective behavior 
including pairing, density modulations, and emergent SU(N) symmetries in 
optical lattices, while establishing whether an elementary scattering framework 
captures the crossover from universal s-wave physics to anisotropic or even 
topological many-body phases. Resolving these points would establish field-linked 
molecules as a fully tunable building block for designer quantum matter.

\subsection*{Many-body probes, superfluids-solid interfaces and quantum groups} 

We are also building theoretical frameworks for impurity-based quantum sensing 
as a probe of many-body systems. Localized impurities, being atomic, molecular, 
or solid-state, serve as tunable spectrometers of low-energy excitations, providing 
order-parameter–agnostic signatures of quantum phase transitions via shifts in 
dressing, linewidths, and non-linear response. 

In parallel, we are pursuing a complementary theoretical direction based on 
quantum-group methods. 
Building on our expertise in minimal models, we aim to use q-deformed algebras 
to construct tractable yet flexible descriptions of quantum many-body systems, 
providing symmetry-based tools that extend beyond conventional Lie-algebraic 
frameworks.

Other efforts target how superfluids behave at interfaces with structured media. 
Superfluid–solid hybrids provide a natural setting to study momentum transfer, 
emergent friction, and the onset of supersolidity. 
These systems offer a minimal framework to understand how microscopic interfacial 
dynamics give rise to macroscopic transport and collective phenomena, shedding light 
on how solid structure can reshape superfluid responses.

\noindent
\vspace{1cm}

\section*{Advanced Cluster Experiments}
\label{F{\'a}rn{\'i}k}
\noindent
Michal F\'{a}rn\'{i}k\\

\textbf{Two experiments are proposed to promote the molecular beam experiments with clusters 
to a next level in relevance to atmosphere, as well as to provide more understanding of
elementary processes in complex molecular systems.}

The introduction of crossed molecular beams to study reactions between two 
individual molecules marked a significant step toward understanding chemistry at a 
detailed molecular level. 
Reactions can change significantly in a solvent environment, which is indispensable 
in chemistry. 
Clusters in molecular beams allowed the investigation of elementary processes in a 
solvent while still providing molecular-level insight. 
Ideally, to understand the evolution of a certain process from individual molecules 
to a bulk, we would like to investigate it as a function of the cluster size, adding 
the molecules to the system one by one. 
This can be achieved in calculations or experiments with charged clusters, which can 
be selected by mass spectrometric methods, but \textit{size selection of neutral clusters} 
is difficult.

Some size selection methods were reported for relatively small clusters: elastic 
scattering with a secondary beam of helium atoms~\cite{Buck84,Buck14persp} and methods 
based on deflection of the cluster beam in an electrostatic deflector for clusters 
with dipole moments~\cite{Trippel2012}. 
Neutralization of mass-selected cationic cluster ions by charge transfer can also produce 
size-selected neutrals~\cite{Arnold1985}. 

\begin{figure}[h]
\centering
\includegraphics[width=0.9\textwidth]{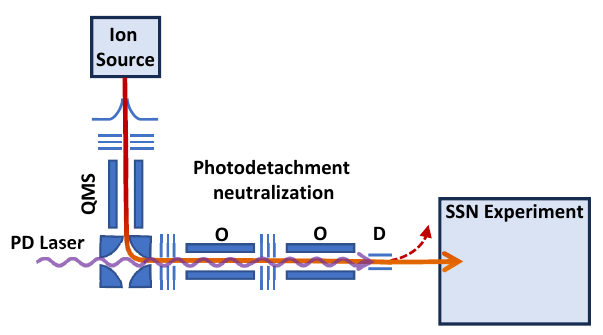}
\caption{Sketch of size-selected neutral (SSN) cluster experiment based on photodetachment 
of electrons from mass selected negatively charged clusters}\label{f:exp}
\end{figure}

We propose photodetachment (PD) of electrons from mass-selected anion clusters to 
generate a beam of neutral clusters of a selected size. 
Photoelectron spectroscopy of negatively charged clusters is a mature field, and 
many different anionic clusters were investigated~\cite{Wang2017,Newmark2001}. 
Absorption cross sections for PD are typically relatively large~\cite{Luzon2016}. 
Thus, overlapping a laser beam with the size-selected negative ions on a long enough 
absorption path can produce a beam of the corresponding neutrals sufficiently intense 
for further experiments. 
The proposed setup is sketched schematically in Fig.~\ref{f:exp}. 

Different types of ion sources may be used to produce negatively charged clusters, 
e.g., electrospray or supersonic expansions combined with electron attachment. 
A higher temperature resulting in the metastability of negatively charged clusters 
~\cite{Poterya24JPCA_water} from such sources could be of concern here. 
However, cluster anions might be cooled in supersonic expansions like in magnetrons 
or combined expansions with laser-generated microplasmas. 
Anions are stabilized by collisions in such sources as recently observed for 
naphthalene cluster anions~\cite{Durana2024}. 
An ion trap can be added, where the anions are cooled by collisions with cold gas 
molecules before they are extracted to the PD region. 
The neutral clusters can still fragment after photodetachment because of the difference 
in geometry between the anionic and neutral species. 
However, this effect can be suppressed in larger clusters - our species of interest 
here - by distributing the excess energy among their many degrees of freedom.         

The ions can be mass selected by a quadrupole and bent to overlap with the laser on 
a long flight path, on which the ions are contained by octupole ion guides. 
Afterwards, the remaining ions are deflected from the beam, while the neutral size 
selected clusters can be used for experiments further downstream. 
They can interact with electrons or/and photons of different energies in analogy to 
our experiments performed routinely with neutral clusters 
~\cite{Farnik18MSR,Farnik21PCCPersp,Farnik23Persp}. 
However, detailed size-specific information will be gained, which is often smeared 
out in experiments with a broad neutral cluster size distribution.


The second proposed experiment was partially inspired by one of the talks at the 
Erice meeting. 
There is interest in \textit{collisions between clusters} in the astronomic community, 
since the processes that occur upon collisions between cosmic ice/dust grains are very 
important for the chemistry and physics in molecular clouds~\cite{Papoular2004}. 
In the atmosphere, cluster-cluster collisions are also important: proton transfer 
between colliding graupel particles in thunderstorm clouds is assumed to be responsible 
for cloud charging and lightning~\cite{Jungwirth2005AR,Williams2005}. 
Such processes are hardly studied in laboratory experiments, and our knowledge about 
them stems mostly from field observations complemented by theoretical calculations. 

Collisions between individual molecules were studied in crossed molecular beams, as 
well as collisions of atoms and molecules with clusters. 
Collisions between two clusters could also be investigated in crossed-beam experiments. 
The analysis of such scattering experiments will be complicated by the neutral 
cluster size distributions in both colliding beams.
However, for some species, the size distributions are known functions of the 
expansion conditions. 
Another possibility is to use charged, and thus mass-selected, clusters in one of the 
colliding beams, in analogy to the crossed beam ion-molecule collision experiments. 
This also provides the advantage of tuning the collision energy by the acceleration 
voltages. 
The major concern in these experiments would be the beam densities to achieve enough 
cluster-cluster collisions.

The two above-proposed experiments could elevate the cluster beam experiments to 
the next hitherto unexplored level, with relevance for atmospheric chemistry and 
astrochemistry as well as for various fundamental processes.

\subsection*{Acknowledgement}
This work is supported by the Czech Science Foundation project 24-11390S.

\noindent
\vspace{1cm}

\section*{Two-dimensional electron energy loss spectroscopy of clusters}
\label{Fedor}
\noindent
Juraj Fedor\\

\textbf{It is proposed that a trochoidal electron spectrometer is utilized for 
the purpose of vibrational electron energy loss spectroscopy of clusters. 
This will enable probing the effects of aggregation on dynamics of resonances.}

Formation of electronic resonances (temporary anions) in electron-molecule collisions 
represents an elegant way to probe ultrafast nuclear dynamics. 
The trick lies in monitoring the energies of the autodetached electrons. 
We can denote the scattering process as
\begin{equation}
e^- (E_{\rm i}) + {\rm AB} \to {\rm AB}^{-\#} \to {\rm AB}(E_{\rm vib}) + e^-(E_{\rm out}),
\label{eq:eels}
\end{equation}
with $E_{\rm i}$ and $E_{\rm out}$ being the energies of incident and outgoing 
electrons, respectively. 
Their difference, $E_{\rm vib}$ is left as an internal energy of the target (here we will 
consider only excitations of the nuclear degrees of freedom, hence the subscript). 
AB$^{-\#}$ stands for the resonant state. 
The dynamics of this state influences $E_{\rm out}$: 
in the simple adiabatic picture, the closer the potential energy surface of 
AB$^{-\#}$ to that of AB in the geometry of the (vertical) detachment, the 
smaller the $E_{\rm out}$. 
However, it is well known that the adiabatic picture very often fails and the 
dynamics is strongly non-Born-Oppenheimer and shows nonlocal effects (non-locality 
is related to the fact that an electron which detaches at a certain nuclear 
configuration can re-attach when nuclei are in different positions).  

Such scattering can be probed by two-dimensional electron energy loss spectroscopy 
(EELS), where $E_{\rm i}$ is controlled, outgoing electrons are energy-analyzed, and 
their signal is plotted as a color map in ($E_{\rm i}, E_{\rm out}$) space. 
In the recent decade, this approach has revealed a number of intriguing dynamical 
effects on resonances, e.g., distant symmetry control of molecular bond 
cleavage~\cite{Kumar:Pyrrole:2022}, effects of vibronic coupling in 
continuum~\cite{Dvorak2022PRL} or selectivity in vibrational autodetachment 
following complete randomization~\cite{anstoter_NB20}.  
The natural question arises as to how these phenomena change upon clustering. 
Apart from the expected effect of increasing the number of degrees of freedom 
and the question of how this influences non-ergodic vs. ergodic behavior, one can 
speculate about the possible influence of a change of electron affinity, dipole moment, 
polarizability, etc. 
All of these are known to have profound influence on resonant 
scattering~\cite{Fabrikant_JPB_2016}.

The interest in the resonance dynamics in clusters is fostered by the results of 
photodetachment experiments. An alternative way to populate a resonance is to 
photoexcite a bound anion:
\begin{equation}
\hbar \omega + {\rm AB}^- \to {\rm AB}^{-\#} \to {\rm AB} (E_{\rm vib}) + e^-(E_{\rm out}).
\end{equation}
Such experiments can also be performed in a 2D manner with varying wavelengths of 
light where the signal of detached electrons is color-mapped in the 
($\hbar \omega, E_{\rm out}$) space (2D photoelectron spectroscopy, PES)~\cite{Anstoter16}. 
Since neutral AB and stable anion AB$^-$ often have a different equilibrium structure, 
applying 2D EELS and 2D PES in the same molecule reveals the effect of the starting 
geometry on the resonance dynamics~\cite{rankovic_NB22}.

There is, however, a strong imbalance: while photoelectron spectroscopy has been 
extensively applied to cluster anions, the EELS experiments on clusters are basically 
completely absent. 
The reason for the latter fact is the sensitivity of the typical EELS arrangement, 
where the incident beam is monochromatized, and
the electrons scattered into a narrow angular range are analyzed, typically by an 
electron analyzer which passes only a narrow range of $E_{\rm out}$ at a time (e.g, 
a hemispherical analyzer). 
Thus, even for gas-phase targets for which measurements are done very close 
($\approx 1$ mm) to the effusive nozzle, acquisition of one 2D EELS is extremely 
time consuming (typically several weeks of net acquisition time). 
A neutral cluster beam resulting from an expansion typically has a much lower 
local density, which makes the prospects of 2D EELS of clusters grave. 
Indeed, there are only very few one-dimensional EEL spectra of rare-gas clusters 
available in the literature~\cite{burose91, allan93}.
On the other hand, the cluster photodetachment experiments are facilitated by 
the high photon flux of the laser light, by the possibility to use velocity map imaging 
for $E_{\rm out}$-analysis and by using ion traps to increase the local density of 
AB$^-$ available to probe anionic clusters. 

Is there a possible solution to probe the process~(\ref{eq:eels}) in clusters? 
There is not much one can do about the density of the molecular cluster beam, 
apart from having the interaction spot as close to the nozzle/skimmer as possible. 
However, there might be a way to increase the experimental sensitivity by using 
a different type of electron monochromator and analyzer. 
The idea is to use a trochoidal electron energy loss spectrometer. 
It utilizes one $\vec{E} \times \vec{B}$ dispersive element to define the 
incident beam and two such elements to analyze the scattered electrons. 
Due to the collimating magnetic field and the fact that electrons scattered 
into a broader range of scattering angles are collected, it has a superb 
sensitivity. 
This is documented by a large amount of one-dimensional EEL spectra with excellent 
signal-to-noise ratio~\cite{allan_habil89}. 
At the same time, the typical resolution is around 30-50~meV, sufficient to 
resolve the main groups of excited vibrations. 
The trochoidal EELS was invented in 1989 by M. Allan~\cite{allan_habil89} 
and was adopted by only several laboratories, e.g, in Pittsburgh~\cite{falcetta91} 
or Belgrade~\cite{vicic98}. 
What prevented its wider spread was probably the difficulty in machining the 
electron optics which requires high precision, especially in aligning the orifices. 
However, current high-precision machining technology should enable routine 
construction of trochoidal spectrometers and their combination with cluster beams. 

\subsection*{Acknowledgement}
This work is supported by the Czech Science Foundation project 24-11166S.

\noindent
\vspace{1cm}

\section*{Probing the astrochemical network of transition metal sulfides with 
gas-phase infrared spectroscopy}
\label{Ferrari}
\noindent
Piero Ferrari, Laura X. Worutowicz, Rick J. Louwerse, Denis Kiawi,
Laurens B. F. M. Waters, Sandra M. Lang, Joost M. Bakker\\

\textbf{The formation of iron sulfide clusters could provide a possible sink for 
the significant depletion of atomic sulfur in diffuse to denser regions of the 
interstellar medium. In order to explore this possibility, spectroscopic information 
about the clusters is required, ideally under the cold and isolated conditions of 
molecular beams. 
Here, we discuss the possible routes to form gas-phase Fe$_n$S$_m$$^+$ clusters. 
In addition, we detail the need for broadband spectroscopic information, from the 
far-infrared to the near-UV. This information, in combination with astronomical 
observations, could provide key pieces of information to solve the sulfur depletion 
problem. 
Analysis of the electronic structure of the clusters highlight the possibility that 
iron sulfides act as strong recurrent fluorescence emitters, making them stable 
species in the harsh conditions of the interstellar medium.}

Clusters are known as an intermediate form of matter linking the atomic and bulk phases. 
Similarly, they could form a bridge between different forms of matter in the Universe, 
from the atomic species generated in nucleosynthesis to the solid forms in interstellar 
and circumstellar matter. 
At such, gas-phase cluster science can provide the level of detail required to understand 
the formation of solids in space. Clearly, the idea is not novel. 
Already after the discovery of C$_{60}$, the possibility that this large molecule is 
present in the interstellar medium (ISM) was speculated~\cite{KrotoNature1985}, 
something which was soon after confirmed by careful spectroscopic 
measurements~\cite{cami2010detection}. 
With this idea in mind, a wealth of cluster systems has since been investigated, 
e.g. alumina, silicates and met-cars~\cite{de2024gas,demyk2004experimental,castleman2009clusters}. 
Here, we showcase another potentially abundant species in space, iron sulfide clusters.
\begin{figure}
    \centering
    \includegraphics[width=0.6\linewidth]{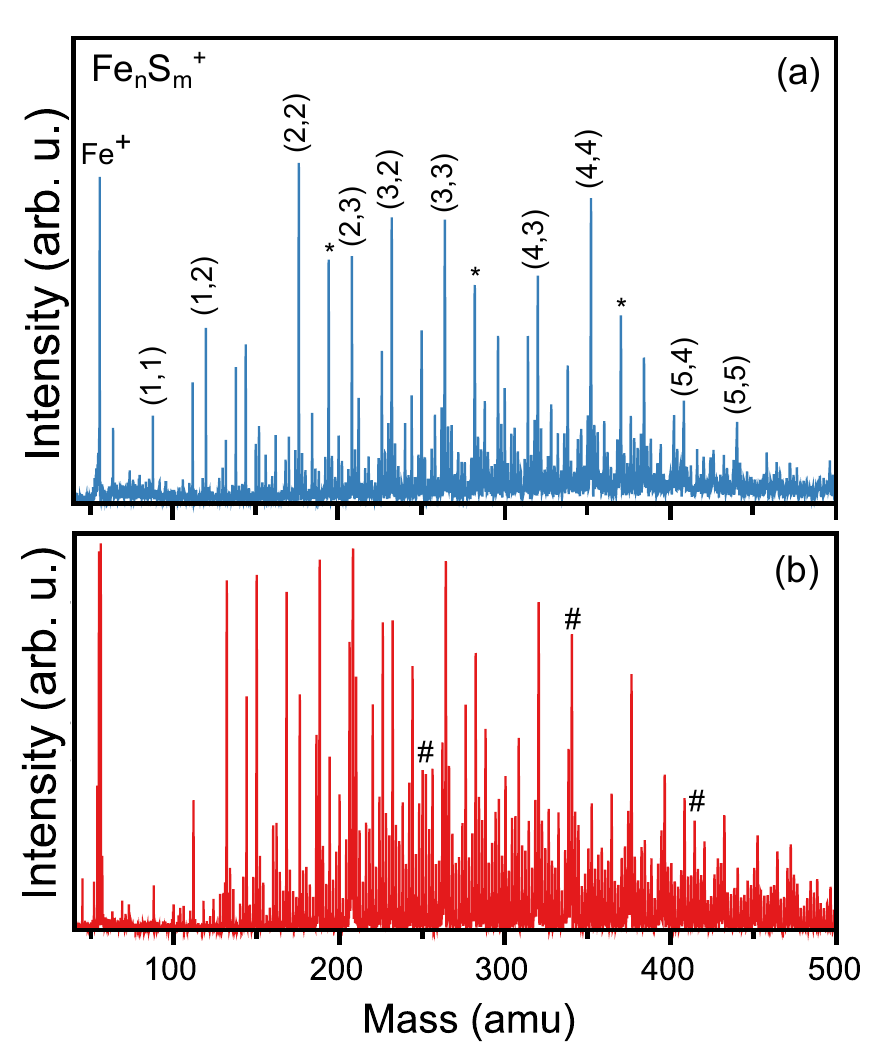}
    \caption{Examples of a mass spectra of Fe$_n$S$_m$$^+$ clusters, produced 
		by the laser ablation of pyrite (a) or by seeding CS$_2$ into the carrier gas (b). 
		Selected peaks are labeled based on the ($n$,$m$) composition. 
		Some complexes with CS$_2$ attached are marked by the $\#$ symbol, whereas 
		H$_2$O contaminants are highlighted by asterisks.}
    \label{MassSpectrum}
\end{figure}
Regarded as one of the six life-essential chemical elements of life as we know it, 
understanding the chemical evolution of sulfur in the ISM remains enigmatic in 
astrochemistry. 
While observations of diffuse interstellar environments show a sulfur abundance 
roughly corresponding to the expected cosmic value~\cite{goicoechea2006low}, observations 
of dense molecular clouds, star-forming regions and planet forming disks reveal a 
major depletion of the amount of gas-phase atomic sulfur, by up to two orders of 
magnitude~\cite{le2021molecules}. 
Still, sulfur is found in rocky sources~\cite{calmonte2016sulphur}.

One solution to the sulfur depletion problem could be the formation of sulfur-bearing 
molecules in denser interstellar environments, such as HCS, OCS, SO$_2$ and NH$_4$SH. 
Although these have indeed been observed, it is in amounts too low to explain the 
detected sulfur depletion~\cite{mifsud2021sulfur,van2024joys+,slavicinska2025ammonium}. 
Pure sulfur clusters, like 
S$_8$~\cite{ferrari2024laboratory,shingledecker2020efficient,taillard2025predicting}, 
have also been considered, but their detection still awaits. Another possibility 
yet to be explored is the locking of sulfur in metal sulfides. 
Given the relatively high cosmic abundance of iron, solid FeS is an important candidate, 
which in fact has been found as a sulfur reservoir in meteoritic material~\cite{avril2013raman}. 
In this respect, it is crucial to provide the spectral signatures of these species, 
ideally under the cold and isolated conditions of molecular beams, which resemble well 
those of the ISM.

The first step towards spectral characterization is the formation of 
Fe$_n$S$_m$$^+$ clusters in the gas phase, which was achieved earlier 
using sputtering~\cite{heim2015challenge} and through laser ablation of 
pressed targets of mixed sulfur and iron powders~\cite{lang2018thermal}. 
Laser ablation of pressed targets, however, is notorious for stability issues. 
Consequently, other approaches to form Fe$_n$S$_m$$^+$ clusters have been considered. 
Figure \ref{MassSpectrum}a presents a mass spectrum of Fe$_n$S$_m$$^+$ clusters 
formed by the laser ablation of the mineral pyrite. 
A wide distribution of mixed clusters is formed, with prominent peaks corresponding 
to clusters with stoichiometric ($n$ = $m$) composition, thus different from 
the expected FeS$_2$ ratio of pyrite. 
Many other $n$,$m$ compositions are seen in the mass spectrum, showing the 
richness of the distribution. Still, with this approach, there is not much 
flexibility in the composition of the formed clusters, given the fixed amounts 
of Fe and S atoms in the pyrite target. 
Moreover, ablation of a rocky substrate proved to be also susceptible to 
instabilities.

Another alternative for the production of Fe$_n$S$_m$$^+$ clusters is to 
use a pure iron target for ablation, but seeding a sulfur-containing molecule 
into the He carrier gas. H$_2$S, CS$_2$ and OCS were tested, all yielding much 
richer mass spectra, in which not only pure Fe$_n$S$_m$$^+$ clusters are 
observed, but also complexes of Fe$_n$S$_m$$^+$ with the seed molecule. 
An example is shown in Figure \ref{MassSpectrum}b, using CS$_2$ as seed. 
Crucially, changing the percentage of CS$_2$ or OCS gas in He allows control 
over the composition of the formed clusters. 
We note that the formation of pure Fe$_n$S$_m$$^+$ implies the dissociation 
of H$_2$S, CS$_2$ or OCS, which is likely not exothermic or kinetically 
favorable in the aggregation zone~\cite{koszinowski2004formation}. 
This shows an important role of the ablation process; the relative timings of 
gas injection and laser ablation can therefore provide additional control 
over the composition.

Gas-phase IR spectroscopy has a long history in cluster science, used to 
determine the geometry of isolated clusters~\cite{asmis2012structure}, as 
well as elucidating their chemical activity~\cite{fielicke2023probing}. 
With the commissioning of the James Webb Space Telescope (JWST), covering 
a wavelength range from the visible down to $\sim$ 350 cm$^{-1}$, JWST has 
provided IR spectra of many astrochemical environments with exquisite 
precision~\cite{chown2024pdrs4all}. In the context of the sulfur depletion 
problem, JWST observations have revealed the presence of sulfur-bearing 
molecules in interstellar ices or in star forming regions, such as OCS 
and SO$_2$~\cite{mcclure2023ice}. Clearly, these detections rely on the 
known IR spectra of the molecules, an information often not available for 
complex clusters under cold and isolated conditions.

Therefore, recording the IR spectrum of iron sulfide clusters under gas-phase 
conditions can provide the necessary information to detect such species in 
space, adding key information to solve the long standing sulfur depletion 
problem. Recently, we reported the IR spectra of S$_8$, S$_4$$^+$ and 
S$_4$$^-$~\cite{ferrari2024laboratory}. 
The low frequencies of the vibrational 
modes detected, all below 500 cm$^{-1}$, reveal the necessity to use an IR FEL 
for such studies. Inclusion of Fe will likely lead to similar low frequencies.
\begin{figure}
    \centering
    \includegraphics[width=0.6\linewidth]{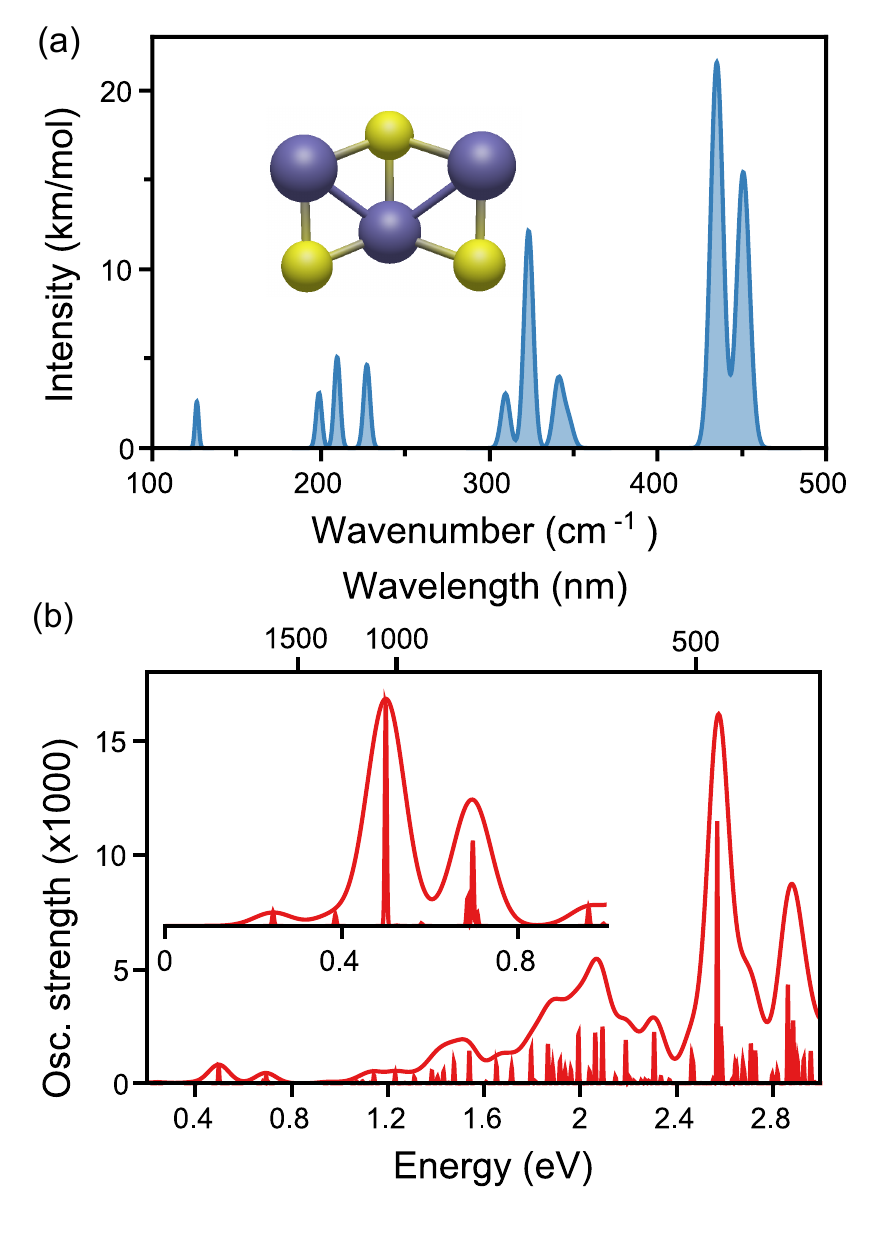}
    \caption{Computed IR (a) and optical (b) spectra of the found lowest-energy 
		structure found of Fe$_3$S$_3$$^+$ (quartet multiplicity), given as oscillator strength
		vs. photon energy. 
		The spectra are broadened by Gaussian line-shapes with a 
		wavenumber-dependent (1$~\%$ of the central frequency, panel a) and a 
		fixed (0.1 eV, panel b) linewidth, for visualization purposes. 
		Vibrational frequencies (panel a) are scaled by a factor of 0.98.}
    \label{Spectra}
\end{figure}

Figure \ref{Spectra}a presents a density functional theory (DFT) calculation 
(PBE/Def2-TZVPP using ORCA 6.1~\cite{neese2025software}) of the harmonic 
vibrational spectrum of Fe$_3$S$_3$$^+$, one of the most abundant species 
observed in the mass spectra. 
The lowest energy structure found for Fe$_3$S$_3$$^+$ 
is shown as an inset, but we stress that many other configurations are possible, 
which emphasizes the necessity of experimental investigations. 
The IR spectrum is rich below 500 cm$^{-1}$, with a strong double feature at 
435 and 450 cm$^{-1}$, both within reach by JWST. At this point, recording the 
IR spectra of the Fe$_n$S$_m$$^+$ clusters becomes pressing. First, this will 
allow for determining the lowest energy configuration of the clusters, which in 
combination with photofragmentation experiments can provide information about 
relative stabilities. 
Moreover, the IR spectra can be combined with JWST 
observations in order to identify iron sulfide clusters in space. 

While the experimental IR spectra of Fe$_n$S$_m$$^+$ clusters are 
significant, we stress that other wavelengths are also relevant and worth 
investigations, for example 
in the context of the diffuse interstellar bands (DIBs). Using again 
Fe$_3$S$_3$$^+$ as an example, 
a time-dependent DFT (TD-DFT) calculation of the optical absorption spectrum 
of the cluster (Figure \ref{Spectra}b) reveals that Fe$_3$S$_3$$^+$ is optically 
active in the visible and near IR 
wavelength region. 
Many predicted electronic transitions between 400 and 1000 nm (3.10 and 1.24~eV), 
in particular the 
stronger features at 430 nm (2.88 eV), 482 nm (2.57 eV) and around 620 nm (2.00 eV), 
fall into the range of the DIBs. 
Broadband spectroscopic studies from IR to visible and near-UV are therefore of 
interest to characterize iron sulfide clusters. 

An interesting feature in Figure \ref{Spectra}b is the prediction of electronic 
states lower than 1 eV, well below the computed fragmentation energy of 4 eV for 
Fe$_3$S$_3$$^+$. 
In particular, the calculation shows transitions at 0.70, 0.50 and 0.38 and 0.25 eV. 
The presence of such states could indicate an active recurrent fluorescence (RF) 
channel in iron sulfide clusters~\cite{FerrariIRPC2019}, providing a fast 
de-excitation mechanism and thus an efficient stabilization factor for isolated 
species present in energetic interstellar environments. 
Many carbon-bearing molecules~\cite{ebara2016detection,stockett2023efficient} 
as well as some metal clusters~\cite{PeetersPRR2021,hansen2017thermal} 
are nowadays known to be RF emitters. An active RF cooling channel would have 
significant astrochemical implications so it deserves further investigation. 
Ideally, direct spectroscopic information about these low-lying electronic 
excitations would be obtained, as recently done for Co$_n$$^+$ clusters using 
FELIX~\cite{kaw2024direct}. Moreover, the cooling dynamics of excited 
Fe$_n$S$_m$$^+$ clusters can be studied on storage ring facilities or even 
using time-of-flight mass spectrometers, depending on how fast RF is on iron 
sulfide clusters.

\subsection*{Acknowledgement}
We gratefully acknowledge the Nederlandse Organisatie voor Wetenschappelijk 
Onderzoek (NWO) for the support of HFML-FELIX and for CPU time on the
Dutch National Supercomputer Snellius (Project 2021.055). 
This work was possible thanks to the constant support of the FELIX staff.

\noindent
\vspace{1cm}

\section*{The elusive parameters of unimolecular decays}
\label{Hansen}
\noindent
Klavs Hansen\\

\textbf{The decomposition of a single molecule driven by its internal 
excitation energy, known as a unimolecular reaction, is particularly 
relevant for molecular beams, the birthplace and still a favored venue 
for their study.
This contribution sketches a concrete experimental protocol that can be 
used to determine the important and otherwise unreachable parameters in cluster 
decays.}

\subsection*{Background}

The description of thermal decay of isolated molecules has been an intensely studied 
subject for more than a century.
The interest in the subject is witnessed by the number of theories describing these
reactions (RRK, RRKM, various phase space theories, detailed balance), but by the 
same token the multitude also demonstrates the lack of consensus on the
ultimate understanding of the processes.

Apart from the desire to have a fundamental understanding of these decay processes,
the questions of determining the parameters describing unimolecular reactions is a very 
practical one for clusters. 
The reactions are frequently used as the action part of action spectroscopy. 
Often the observed action is equated with absorption, which requires a quantum 
yield of unity.
At long wavelengths and/or for large clusters this will invariable introduce a spurious
spectral cutoff.

A first attempt to determine parameters of the rate constants encounters a severe 
problem from the range of excitation energy over which one can observe 
unimolecular decays.
Experimental time ranges combined with the strong variation of the rate constants 
severely limit the energy range feasible for measurement. 
This makes fits of the parameters numerically problematic, even in 
the best cases.
Best cases are provided by experiments at ion storage rings 
~\cite{Moeller_1997,Jinno04,Schmidt_2013,Nakano_2017,Meyer17} and similar devices 
~\cite{ZajfmanNIMA2004}.
A few of these devices operate under cryogenic conditions, providing the very 
low pressure needed for long term ion storage, at present beyond thousands of 
seconds.
However, the quenching effect of thermal radiative cooling imposes its own time 
cut-off on millisecond times scales by IR radiation~\cite{NajafianJCP2014,IidaPRA2021,
HansenPRA2017,Stockett2019}, and can even be active at timescale as short as 
microseconds or tens of microseconds by the emission of radiation 
from thermally excited electronic states~\cite{HansenJCP1996,andersen1996,
ItoPRL2014,HansenPRA2017}.

The confrontation of the multitude of theories with experiments are particularly 
difficult for clusters.
One of the reasons for this is that clusters are intrinsically thermodynamically 
unstable, even though they are usually stable on an absolute scale.
This leaves the molecular beam (traps, etc) as the only venues where the 
free cluster, the pristine cluster, can be studied.
Such beams come with a lack of contact with a thermalizing heat bath, which rules 
out constant temperature measurements.

An alternative to a thermal equilibrium experiment is to measure decay 
rates for clusters at a specific energy.
As a rule of thumb, for a thermally activated process that proceeds on time scales of 
tens of microseconds, a change of the excitation energy of 2~\% will change the rate 
constant a factor of two.
To generate clusters with excitation energy distributions below this limit
comes with its own problems.
Although thermalization of clusters to low and narrow energy distributions are 
possible and have been accomplished, as demonstrated in 
ref.~\cite{IidaPRA2023}, large heat capacity systems are excluded from the technique 
used in that reference. 
The energies of the photons needed in the second step of the procedure
grow with the size of the system and ionization will prevent the necessary soft energy transfer. 
Other techniques often suffer from uncontrolled energy distributions.
An example is found in~\cite{YeretzianScience1993}. 
Interestingly, similar problems can arise for reactivities measured in a apparently 
thermalized gas phase reaction chamber~\cite{Prigogine1958}.

The precise parameters that are relevant for an exact description of the reaction
depend on which of the many theories from the literature one adapts, but as
a robust parametrization one can represent a rate constant as the product of a 
frequency factor, an activation energy 
and a temperature, the latter two appearing as a ratio in the argument of 
an exponential, i.e., as a simple Arrhenius expression.
In the absence of a sufficient dynamic range of energies one often resorts to fitting 
models with predetermined values for some of the parameters decided by
one's favorite theory.
The usual procedure is to calculate theoretically the frequency factor and the caloric 
curve and use the data to fit the activation energy of the process.
This is a risky proposition for rate constants; any error in the preconceived 
frequency factor can be masked by a corresponding incorrect value of the activation 
energy.
The frequent presence of anharmonicities in the vibrations cause other problems.
The vibrational density of states is needed to convert an energy to the effective 
temperature, and anharmonicities introduce an uncertainty in this conversion which 
goes beyond a perturbative correction.

We seem to be up against a wall of obstacles.

\subsection*{Proposal}

The experiments proposed here circumvent the problem with the broad energy distributions,
by turning the obstacle they present into an advantage.
The key observation is that under fairly general conditions the broad 
excitation energy distributions cause 
decay rates that vary with time as $1/t$, where $t$ is the time since 
production~\cite{HansenMSR2021}.
This claim is borne out by experiments~\cite{HansenPRL2001}.
It is clearly manifested in decays of molecules and clusters in electrostatic ion storage 
rings, in which decay rates are one of the primary measured signals, and has paradigm 
status for these types of experiments.

The expression for the rate, $R$ is, with all the factors given 
explicitly~\cite{HansenMSR2021}, 
\begin{equation}\label{rate1}
R(t) = cg \frac{E_{\rm a}C_{\rm v}}{k_{\rm B}\ln(\omega t)^2 t},
\end{equation}
where $g$ is the density of excitation energy in the ion ensemble, for simplicity 
assumed constant here, 
$E_{\rm a}$ is the reaction activation energy, $C_{\rm v}$ the average heat capacity of
the reactant and product, $k_{\rm B}$ is Boltzmann's constant, $\omega$ is the 
rate constant frequency factor, and $c$ the transmission and detection efficiency.

In contrast to an exponential decay, this rate comes without a time scale
and therefore neither with an implied energy scale.
Absorption of a single photon at a well defined time, $t_{\rm las}$, will
provide an energy scale.
An absorbing cluster will be reheated to produce a decay rate corresponding 
to the decay rate that occurred at an earlier time, $t_0 < t_{\rm las}$:
\begin{equation}\label{rate2}
R_{\rm las}(t) = p R(t- t_{\rm las} - t_0) + (1-p)R(t), 
\end{equation} 
where $p$ is the photon absorption probability.
The backshifted rate $R(t-t_{\rm las})$ can be found from this expression,
with the cross section as a bonus.
The connection to the rate constant, $k$, appears with the insight that the decay at any 
given time occurs from clusters that have rate constants centered on the 
value $1/t$~\cite{HansenMSR2021},
\begin{equation}
R(t) \appropto k(t) \propto 1/t.
\end{equation}
Up to a constant we can therefore identify the rate constant at $t_0$ and $t_{\rm las}$
with $1/t_0$ and $1/t_{\rm las}$ or, more generally, with $R(t_0)$ and $R(t_{\rm las})$.
Hence,
\begin{eqnarray}\label{Randk}
R(t_{\rm las}) &\propto& k(E(t_{\rm las})),\\\nonumber
R(t_0) &\propto& k(E(t_0)) = k(E(t_{\rm las})+h\nu),
\end{eqnarray}
where $h\nu$ is the photon energy.
This determines the absolute value of the energy loss between $t_0$ and $t_{\rm las}$
to $h\nu$.
With the insight from Eq.~(\ref{Randk}) and a sufficient number of measurements with 
different photon energies and different laser firing times, we can link rate constants 
for different energies, and a fit of the parameters in the expression 
\begin{equation}
k(E) = \omega \exp\left(- E_{\rm a}C_v/k_{\rm B}(E+E') \right)
\end{equation}
is in principle possible.
Here $E$ is the excitation energy, $C_v$ is the effective heat capacity, and $E'$ the
offset one expects in the canonical caloric curve, $E = k_{\rm B} T - E'$.

An analysis of this kind was already performed on the data recorded for the decay
\begin{equation}\label{C60e}
{\rm C}_{60}^- \rightarrow {\rm C}_{60} + e^-,
\end{equation}
with the data from~\cite{SundenPRL2009} and published in~\cite{HansenPRA2020}.
It acts as a proof of principle, as the parameters of the molecule
were already fairly well established prior to the experiment, and 
the reaction in Eq.~(\ref{C60e}) is fairly well understood in terms of detailed balance
theory.
A major question is the description of the loss of molecular fragments, specifically 
the role of the rotational entropy of the fragment, which is of particular relevance 
for clusters of molecules and fullerenes~\cite{HansenPRL1997,HansenIJMS2006}.
Experiments based on the present proposal should begin to shed light on the
matter in the hopefully near future.

\noindent
\vspace{1cm}

\section*{Some remarks on future cluster research}
\label{Issendorff}
\noindent
Bernd v. Issendorf\\

\textbf{Cluster physics has come a long way since the first observation of a cluster 
mass spectrum (of carbon dioxide cluster cations) in 1961~\cite{henkes_notizenionisierung_1961}. 
Only six years later the ionization potentials of small sodium clusters were 
measured~\cite{robbins_ionization_1967}, and already then one of the classical 
questions of cluster physics was posed, which still is not fully answered yet: at which 
size can a metal cluster be considered a metal?}

Since then a plethora of experimental and theoretical tools have been developed, and 
applied to a large number of cluster systems, from weakly bound rare gas and molecular 
clusters to covalently bound insulator, semiconductor and metal clusters, addressing 
a broad range of phenomena~\cite{haberland_clusters_1994,sattler_clusters_2010}.

Many of these studies concentrated on the cluster structure, as its knowledge is 
indispensable for a detailed understanding of any cluster 
property~\cite{baletto_structural_2005}. 
Cluster structures have been determined by the combination of spectroscopic measurements 
and calculations. 
Experimental methods like mobility measurements, electron diffraction, infrared 
absorption spectroscopy and photoelectron spectroscopy have seen continuous improvements 
due to the development of better spectrometers and, notably, to the introduction of 
methods to cool the clusters to cryogenic temperatures~\cite{GerlichACP1992}. 
On the theory side, both quantum mechanical methods, mostly DFT, and optimization 
algorithms have also been continuously improved. 
Here of course the enormous development of available computer power was very helpful. 
Newest developments are the introduction of machine learning interatomic potentials, 
which again speed up the calculations a lot~\cite{wang_accelerated_2022}. 
No doubt soon also AI methods for finding cluster structures will be introduced; 
so despite the fact that there are still countless unassigned cluster structures 
this can be considered an almost finished task.

Many other cluster properties have been studied in detail as well, like photo 
absorption spectra, magnetic and chemical properties, or phase transitions like 
the melting transition~\cite{schmidt_irregular_1998,aguado_melting_2011}. 
Quite a number of studies were also devoted to the decay of electronically 
excited states~\cite{koop_long-lived_2017}.

So what are the questions which have not been answered up to now or only partially? 
In the following I will discuss two examples.

\subsection*{Photoionization}

Photoionization of bulk matter seems to be well understood since Einstein's 
interpretation of the photoeffect~\cite{einstein_uber_1905}. 
In fact, this is not the case, at least not in the sense that it can be treated 
easily in ab initio calculations. 
The main reason is that photoionization involves the conversion of a quasiparticle 
into a real particle. 
The quasiparticles, excitations of the Fermi liquid, behave like noninteracting 
electrons, but in fact result from the strong interaction among the electrons as 
well as between the electrons and the background ions, which, for example, alters 
their effective mass.

Photoemission therefore is an involved many-body problem, which should be treated 
on a much higher level than the standard "three step" or "one step" 
approach~\cite{hufner_photoelectron_2003}. 
Clusters should be ideal model systems for this problem, as they can be bulk-like, 
but with a finite (and tunable) number of interacting electrons. 
It has been demonstrated that the angular distribution of photoelectrons emitted 
from simple metal clusters is a very sensitive test for many-body 
effects~\cite{piechaczek_decoherence-induced_2021}. 
Also differential detachment cross sections for photoemission out of different 
initial states should be sensitive tests for theory, especially if the emission 
proceeds via autoionizing excited 
states~\cite{reiners_plasmon_1996,huang_high-resolution_2014}. 
Further work, which should include both high level calculations as well as refined 
experiments, could therefore lead to a better understanding of the photoeffect 
itself, one of the fundamental processes of atomic, molecular and solid-state physics.

\subsection*{Superconductivity}

Superconductivity is one of the most fascinating macroscopic quantum phenomena, 
resulting from the formation of Cooper pairs, weakly bound bosonic pairs of electrons. 
It is one of the oldest questions of cluster and nanoparticle physics how small 
a particle can be and still exhibit superconductivity, or, more precisely, 
Cooper pair formation. 
From the theory side there seems to exist no fully conclusive 
answer~\cite{von_delft_superconductivity_2001,BoseRPP2014}. 
Experimentally Cooper pairing has been observed down to sizes of about 
5 nm~\cite{BoseRPP2014}; in much smaller clusters some exotic behavior has 
been seen, which might or might not be connected to 
superconductivity~\cite{moro_spin_2004,CaoJSNM2008}. 
Given that Cooper pairing in nuclei is a common 
phenomenon~\cite{BMPR1958,dussel_cooper_2007}, 
that is in systems with a rather small number of constituents, one could 
speculate that it might exist also in clusters with just a few ten atoms. 
It is nevertheless unclear how this could be detected in a gas phase and 
therefore unperturbed cluster; theory is not very helpful in this respect as 
there is no calculation yet which takes the (in principle known) electronic 
and vibrational density of states of a real cluster into account. 
Therefore, more advanced experimental and theoretical techniques have to be 
developed to tackle this problem, making it the most challenging and at the 
same time most interesting problem of cluster physics.

Of course, there are many other open questions relating to cluster properties 
like magnetism, chemical reactivity or thermodynamic behavior. 
One could therefore state that there are very interesting times ahead for 
cluster physics.

\noindent
\vspace{1cm}

\section*{Laser-matter interactions in the intermediate regime: 
charting a path through “no-man’s land”}
\label{Kong}
\noindent
Wei Kong\\

\textbf{The field of laser-matter interactions in the intermediate intensity regime, 
called "no man’s land", has resisted a complete description because it lies between 
two theories: quantum mechanics and classical electrodynamics. 
Although several teams worldwide have reported observing multiply charged atomic ions, 
the absence of a theoretical model for the very long evolution times, spanning several 
nanoseconds, and the large number of particles -- tens of thousands of nuclei and 
electrons -- makes the problem seem unsolvable. 
We have begun this work by greatly improving the time-of-flight technology through 
integrating Inverse Problem Theory, gaining much more detailed information on how 
particles are produced and how their velocities change over time and space. 
This detailed data can hopefully stimulate advances in theoretical models, including 
analytical, numerical, and statistical approaches. Ultimately, we aim to develop a 
unified theoretical understanding of laser-matter interactions, covering weak to 
strong fields, from femtoseconds to nanoseconds and microseconds.}

The field of laser-matter interactions in the intermediate intensity regime is 
largely a “no man’s land,”~\cite{FennelRMP2010,ReinhardIntro2004} at the intersection 
of quantum mechanics and classical electromagnetism. 
Several research teams worldwide have reported observing multiply charged atomic 
ions (MCAI)~\cite{KongCPL2004,TranPCCP2024,SharmaJCP2006,StukeCPL1981}, which is 
considered impossible based on current strong-field theories. 
Despite its fundamental significance and exciting applications, this area has 
been abandoned twice over the past forty years: reactions at these intensities 
occur on nanosecond timescales, which are too long for current computational 
methods to simulate, and too short for direct time-resolved imaging of the products. 
To achieve a comprehensive understanding of laser-matter interactions across all 
intensity regimes and timescales, significant progress in this intermediate regime 
is needed, requiring more detailed experimental data, along with substantial 
advances in theoretical frameworks and numerical modeling. 

Experimental studies of laser-matter interactions at intermediate intensities have 
focused on charge state distributions and the average kinetic energies (KE) of 
electrons and ions produced in nanosecond laser fields with nanoclusters of atoms 
and molecules~\cite{KongCPL2004,TranPCCP2024,SharmaJCP2006,StukeCPL1981}.
Qualitative reports on the increasing charge states of MCAI at longer wavelengths 
(from 266 nm to 1064 nm), and on the increasing cluster sizes have been 
published~\cite{KongCPL2004,BadaniMSR2017}. 
However, prior to our work~\cite{ZhangJPCL2020}, there was little information 
on the cluster size, no mapping of kinetic energy distributions, and no detailed 
measurements of laser intensities. 
Our group has made significant progress in time-of-flight technology by applying 
Inverse Problem Theory for data analysis~\cite{TranPCCP2024}. 
We have developed the ability to capture explosion processes on the nanosecond 
time scale: we can now measure particle production times with nanosecond accuracy, 
correlate these times between electrons and ions, and with resolved charge states. 
This capability provides a real-time understanding of ionization pathways for all 
ions and electrons during the cluster disintegration process. 

We have also measured the cluster sizes, the laser beam waist, and 
the laser pulse duration 
of the second harmonic of our Nd:YAG laser~\cite{YaoCPL2023,YaoJCP20211}. 
The scaling law for argon clusters
and core-shell clusters with different molecules has been confirmed from mass 
spectrometry~\cite{YaoCPL2023}.
Additionally, we have demonstrated that if the signal dependence can be expressed 
as a simple power function of the laser intensity, volume averaging does not 
impact the final conclusion about the derived exponent~\cite{YaoJCP20211}. 
However, to fully reveal the saturation effects of multi-photon processes, 
intensity-selective scans involving spatial filters and displacement of the 
laser focus ($z$-scan)~\cite{WalkerPRA1998,DoeppnerEPJD2007,KomarPRL2024}, over 
a sufficiently large dynamic range spanning several orders of magnitude, are 
necessary.  

With these experimental conditions clarified, the dependence of charge state 
distributions on laser intensity, cluster size, and cluster composition can 
be examined~\cite{YaoJCP2021}. 
In a laser field of $10^{11} - 10^{12}$ W/cm$^2$, the relative intensity 
ratios of multiply charged atomic ions stay mostly constant, while the relative 
abundance of Ar$^+$ rises sharply with increasing laser intensity. 
Exponential fits of the yields result in a larger exponent of about 5 for Ar$^+$ 
and a smaller exponent of about 3 for MCAI. 
The width of the TOF profile and thus the kinetic energy of Ar$^+$ also grow 
with higher laser intensities, while the width of the MCAI arrival time remains 
steady across the measurement range.  

Doping the argon cluster with molecules that have lower ionization energies, 
such as fluorene (C$_{13}$H$_{10}$) and 1,3,5-trichlorobenzene (C$_6$H$_3$Cl$_3$), 
significantly reduces the charge state distribution of the MCAI of 
argon~\cite{YaoJCP2022}, although singly charged atomic ions from the molecular 
species, like H$^+$, C$^+$, and Cl$^+$, are positively correlated with the number 
of molecules inside an Ar cluster.
Exponential fittings of the yield of Ar$^+$ eventually reach the same value as 
those of MCAI with increasing molecular concentration. 

The KE distribution of cations from the intermediate intensity regime confirms 
the dominance of Coulomb explosion in the cluster disintegration 
process~\cite{TranPCCP2024}. 
By converting the TOF profile into a three-dimensional velocity distribution, we 
find that the KE of MCAI from doped clusters shows a quadratic dependence on the 
charge of the atomic ions, while for neat clusters, the dependence is cubic, largely 
consistent with the insulating sphere model of Coulomb explosion~\cite{SaalmannJPB2006}.
This result is more similar to reports from extreme vacuum ultraviolet (EUV) 
fields~\cite{Krikunova2012} with similar intensities to reports from near-infrared 
(NIR) intense laser fields~\cite{ParkPRL2022}.
However, the charge state distribution from our experiment is the opposite: we 
observe more higher charge state ions than reported in EUV fields, and our charge 
state distribution actually resembles those reported in NIR fields. 

The KE distribution of photoelectrons is more intriguing, containing three 
groups~\cite{TranJPCC2025}.
The fastest group exceeds the known “cutoff” energy by several orders of magnitude, 
reaching hundreds of eV when the ponderomotive energy $U_{\rm p}$ is only tens of meV. 
The second group, related to above-threshold-ionization (ATI), shows the addition of 
up to 8 photons (over 200~$U_{\rm p}$), an “impossible” process for clusters based 
on previous studies in ultrafast intense fields~\cite{SchuttePRL2015}. 
The slowest group — delayed electrons — are ionized about 100 ns after laser excitation, 
approximately 1 mm downstream from the excitation spot. 
These electrons are tentatively attributed to field ionization of near-threshold 
electrons in the expanding nanoplasma after an initial Coulomb explosion; therefore, 
the delay times of these electrons depend on the strength of the extraction field. 

For both electrons and ions, we observe a significant influence of the external electric 
field on how the clusters evolve after laser excitation, and on the charge state of the 
multiply charged atomic ions. 
The electric field can notably increase the charge of atomic ions, with an average kinetic 
energy per charge rising as much as three times~\cite{TranPCCP2024}. 

These unexpected findings reveal that the intermediate intensity regime is not 
merely an extension of strong fields, and new phenomena require detailed investigation. 
In one of our earlier reports~\cite{YaoJPCL2020}, we suggested that the resonance 
of Ar$_n^+$ at 532 nm plays a crucial role in energy absorption and cluster 
ionization after the initial ionization of the first few Ar atoms. 
This implies that molecular dynamics simulations based on classical electromagnetic 
theory might need to incorporate quantum effects in this regime. 
The KE distribution of the ions also indicates surface 
explosion~\cite{KomarPRL2024}, and the formation mechanism of the MCAI may be linked to 
surface ionization of the highly charged cluster. 
The long laser pulse causes continuous acceleration and ionization of electrons near 
the threshold, leading to a highly positively charged cluster. 

This situation contrasts with reports from strong fields where many exciting 
observations have been made~\cite{RaspeJPCL2025}. 
Although the average KE from the MCAI in the intermediate regime is 
lower~\cite{ParkPRL2022}, the average charge state of multiply charged atomic 
ions in this regime is actually higher than that from fs fields, despite the 
field intensity being three orders of magnitude lower. 

Our findings are just the tip of an iceberg: many more detailed studies are 
necessary to fully understand the process. 
The parameter space for the experimental measurements includes laser properties 
like intensity, polarization, wavelength, and pulse duration; variations in 
the samples, ranging from argon clusters to other atomic and molecular clusters 
of different sizes, as well as core-shell structures with specific sizes and 
compositions; and correlations between wavelength and the physical and chemical 
properties of the sample species. 
Product characterization includes electrons, ions, and photons, along with 
the time dependence of these distributions. 
Regarding experimental techniques, TOF methods, imaging of high-energy particles, 
and other advanced detectors with both position and time sensitivities have not 
been fully utilized. 
Additionally, pump-probe methods, which introduce interruptions to the natural 
evolution of the hot nanoplasma after the initial laser excitation, have yet to 
be implemented in this regime.

Although molecular dynamics simulations have been used for experiments in the 
EUV (ultrafast but moderately intense) and NIR (ultrafast and 
intense)~\cite{FennelRMP2010,SchuttePRL2018}, our results highlight the need 
for a coordinated effort in this area of moderately intense nanosecond laser fields. 
This regime represents the final frontier in laser-matter research, and 
significant theoretical advances, as well as the potential incorporation of 
deep learning from molecular dynamics simulations, may be necessary. 
The goal is to develop a mechanistic understanding of processes by creating a 
unified theoretical framework that connects the strong field of classical mechanics 
with the weak field of quantum mechanics. 

Aside from scientific curiosity, this project could lead to a few technological 
applications. 
The preferred production of high charge state MCAI, when optimized with the 
appropriate laser parameters for yields and kinetic energy distributions, could 
result in a new type of ion source for various applications, such as heavy ion 
therapy in cancer treatment. 
Many exciting developments in ultrafast strong fields have been driven by 
potential applications in nuclear fusion, and the added parameter space provided 
by a nanosecond laser, possibly in combination with a femtosecond laser, could 
make this goal even more attainable. 
We might also achieve control of explosions by manipulating the cluster expansion 
process with an additional pulsed electric field or laser field, producing the 
desired ions or photons in the extreme vacuum ultraviolet.

\noindent
\vspace{1cm}

\section*{Superconducting pairing in individual clusters and in cluster-based granular 
materials}
\label{Kresin}
\noindent
Vitaly Kresin\\

\textbf{A perspective on detection of high-temperature superconducting pairing 
in individual nanoclusters with shell structure is presented. 
This is followed by a discussion of the prospects of combining research on 
size-selected clusters with the growing interest in superconducting granular 
materials for use in quantum electronics and computing.}

Nanoclusters are finite many-body systems, hence exploration of collective 
effects is of utmost interest and importance in cluster science. 
Among the rich diversity of such effects in condensed matter physics, 
superconductivity may be the most celebrated. 
Its striking manifestation, its beautiful and profound quantum-mechanical 
origin, its unique present and future applications, and the broadening 
range of materials and temperatures where it is observed, all contribute to 
the never-ending fascination with this phenomenon.

\subsection*{Pairing in free size-selected nanoclusters} 
Although the most familiar mark of superconductivity is the vanishing of 
electrical resistance, this is not the core requirement. 
The essence of the effect is the transition into a new quantum phase 
characterized by the formation of bound electron pairs, accompanied by a  
restructuring of the electronic spectrum. 
In bulk materials with an initially continuous single-particle spectrum the 
consequence is the opening of a gap at the Fermi surface; in isolated finite 
Fermi systems with a discrete spectrum it is a modification of the spacing 
between the highest-occupied and lowest-unoccupied levels and the appearance 
of odd-even effects. 
The earliest-recognized case (put forward only one year after the appearance of BCS 
theory) is that of atomic nuclei~\cite{BMPR1958}. 
Superconducting correlations significantly impact nuclear level structure and 
rotational spectra~\cite{Broglia2013}. 
BCS pairing is also found in finite clouds of ultracold Fermi 
gases~\cite{GiorginiRMP2008,StrinatiPR2018}. 

Consequently, it is constructive to inquire whether a similar transition 
can take place in metal nanoclusters. 
In fact, a prediction has been made that the answer is an emphatic yes, and in 
fact that the pairing strengths and transition temperatures in clusters should 
be dramatically higher than in the corresponding bulk 
materials~\cite{KresinPRB2006,Kresin2021}. 
The origin of this effect lies in the electronic shell structure of metal clusters. 
The high atomic-like degeneracy of the quantized electronic states is propitious 
for pairing because, qualitatively speaking, the strength of the latter is 
proportional to the density of states at the Fermi level. 
At the same time, as opposed to actual atoms, the clusters also contain a bath 
of vibrational quanta, or phonons, which can couple to the electron shells and 
provide the pairing mechanism.

It is worth mentioning that a so-called “Anderson criterion”~\cite{AndersonJPCS1959} 
is frequently invoked to state that pairing 
cannot take place in systems with too few delocalized electrons $N$ to satisfy
$E_{\rm F}/N > \Delta(0)$, where 
$E_{\rm F}$ is the Fermi energy and $\Delta (0)$ is the superconducting energy gap. 
Using bulk parameter estimates this places the critical particle size 
$d_{\rm cr}$ at 4-6 nm~\cite{BoseRPP2014} or $N > 10^3-10^4$. 
However, as often happens with factoids, the nuance behind this statement tends 
to be lost. 
In cluster shell structure the level spacing is not a uniform ladder as in the 
above estimate, but varies depending on the shell occupancy and cluster 
deformation parameter~\cite{ClemmengerPRB1985}. 
Furthermore, as explained above, the energy gap in a cluster can 
become very much larger than in the bulk. 
These factors show that it is realistic for pairing to arise in certain specific 
clusters with only tens to hundreds of delocalized electrons. 
Further theoretical research supported this conclusion (see, e.g., the references 
given in the review~\cite{EdwardsJSNM2022}).

Strong signatures of such an effect have indeed been detected experimentally in 
some free aluminum clusters made up of only several tens of atoms. 
Photoionization spectroscopy~\cite{EdwardsJSNM2022,HalderPRB2015} revealed the 
onset of a strong bunching-up 
of the electron density of states at temperatures of $\gtrsim 100$~K, which is two 
orders of magnitude higher than the bulk $T_{\rm c}$ but is in agreement with 
theoretical predictions. 
Earlier measurements of cluster heat capacities~\cite{CaoJSNM2008} suggested similar 
behavior. 
(It is worth noting that in a finite system the phase transition is not sharp but 
occurs over a finite temperature range.)

Thus there exists encouraging theoretical and experimental evidence 
of a rather exceptional physical phenomenon: superconducting pairing in individual 
finite metal clusters with critical temperatures exceeding the bulk values by 
about two orders of magnitude. 
With optimization of cluster materials, including 
the capability of cluster sources to produce mixed and alloy-like clusters of 
broadly adjustable composition, it should be realistic to identify systems with 
$T_{\rm c}$’s up to room temperature. 
Research in this direction promises interesting 
fundamental discoveries and certainly merits being re-energized. 

Effectual techniques to obtain beams of pure and mixed metal clusters with 
internal temperatures tunable from above room temperature down to the cryogenic 
range are available (see, e.g.,~\cite{MiaoRSI2022} and references therein), therefore 
electronic spectra can be traced as a function of cluster temperature. 
The most direct way to search for an electron pairing transition is by means 
of photoelectron spectroscopy, focusing on the relative positions and strengths 
of the topmost electronic shells. 
Photoelectron spectroscopy in beams is 
typically performed on cluster anions~\cite{IssendorffARPC2005}, but that should 
not negate the effect because it is the total number of delocalized shell electrons 
that should influence the existence of pairing and not the charge state of the 
cluster.

In conversations about cluster superconductivity the possibility of observing 
the Meissner effect naturally tends to come up. 
Standard textbooks derive the result that a perfectly diamagnetic sphere of 
radius $R$ in an external field $B$ acquires a magnetic moment of 
$\mu = (2\pi/\mu_0)R^3B$. 
With $R^3 = (r_{\rm s} a_0)^3(vN)$, where $a_0$ is the Bohr radius, $r_{\rm s}$ 
is the Wigner-Seitz radius parameter, and $v$ the number of delocalized valence 
electrons per atom, 
this translates into a magnetic moment of $0.08vr_{\rm s}^3\mu_{\rm B}$ per atom 
in a $B = 1$~T field, where $\mu_{\rm B}$ is the Bohr magneton. 
For aluminum with $v = 3$ and $r_{\rm s} = 2.07$ this becomes $\approx 2\mu_{\rm B}$ 
per atom, which is detectable~\cite{deHeerWHandbook2011}. 
However, this seemingly optimistic estimate likely hides a critical flaw: 
the London penetration depth $\lambda_{\rm L}$ in superconductors is orders of magnitude 
larger than the cluster diameter. 
For bulk aluminum its theoretical value is 16~nm, and it is even larger in thin 
films~\cite{BoseRPP2014,LopezSST2025}, hence any measurable flux 
expulsion is unlikely to take place. 
Conceivably one also could attempt measurements at temperatures extremely 
close to $T_{\rm c}$, where in the bulk formalism 
$\lambda_{\rm L}(T) = \lambda_{\rm L}(0)[1-(T/T_{\rm c})^4]^{1/2}$, but 
the diamagnetism here may be obscured by the Larmor diamagnetism of the 
delocalized electrons~\cite{KresinPRB1988}. 
On the other hand, it is not unimaginable that the penetration depth could 
be qualitatively modified in clusters with well-defined shell effects; there 
does not seem to have been a theoretical treatment of this question.

A complementary experimental corroboration could in principle come from detecting 
a critical magnetic field, whereby the spectral features putatively assigned to 
the pairing transition would disappear upon the application of a strong enough field. 
For bulk aluminum $H_{\rm c} \approx 0.01$~T. 
However, the critical fields increase both for stronger coupling/higher $T_{\rm c}$ and 
for lower dimensions~\cite{BoseRPP2014}, so one would need to evaluate carefully 
whether generating an $H_{\rm c}$ high enough to suppress pairing in nanoclusters 
is realistic for a beam apparatus.

From the above, it can be concluded that (a) photoelectron spectroscopy is the 
most realistic and promising tool for detecting superconductivity in clusters; 
(b) theoretical analysis of the magnetic susceptibility of superconducting 
nanoclusters with shell structure has the potential to identify novel regularities; 
and (3) the ability of cluster science to produce and study systems with exceptional 
variability of sizes and compositions offers unique pathways for discovering new 
high-temperature superconducting systems, but systematic progress also requires 
a symbiosis between experiment and rigorous theoretical  analysis of finite 
many-body systems.

\subsection*{Size-selected nanocluster deposition for quantum electronics devices}

In addition to gas-phase studies, cluster science also led to the development of 
size-selective nanocluster deposition tools (see, e.g., the review~\cite{MiriglianoAP2021}). 
A heretofore unexplored but productive application of this technique would be 
for a systematic synthesis and study of granular superconducting films. 

Granular aluminum (“grAl”) first drew attention in the 1960s when it was 
observed that granular metallic films displayed both a higher critical temperature 
and higher critical fields than their bulk counterparts~\cite{BoseRPP2014}. 
This attracted a lot of continuing interest, with as yet no final consensus on 
the main underlying mechanism of this phenomenon.

On the applied side, the past decade has seen a resurgence in interest in grAl 
as a highly promising material for high-impedance quantum circuits. 
Using it for inductive elements allows ultra-high impedance and compact footprints, 
enabling better fabrication of devices such as fluxonium qubits, parametric 
amplifiers, and quantum nonlinear circuits (see, e.g.,~\cite{GuptaPRAp2025} and 
references therein). 

However, the poor controllability of grAl fabrication is a challenge. 
Currently, grAl devices are still grown by the method developed in the 1960s 
and early 1970s: deposition of Al vapor in the presence of oxygen, leading 
to the formation of films with an inhomogeneous distribution of nanoparticle 
sizes (typically in the 1-3 nm range), packing order, and oxidation barriers. 
Sample control is limited to indirect correlations with sheet resistance 
measurements, and particle size estimates are mostly based on data from a 
1973 publication~\cite{DeutscherJVST1973}.

By using modern size-selective nanoparticle deposition one would attain much 
more control and tunability. 
The thickness, distribution, and porosity of the monodisperse film can be 
accurately and reproducibly regulated by varying source parameters, mass 
spectrometer settings, particle landing energy, and deposition time. 
Then, in combination with lithographic patterning, elements such as resonators 
and linear and nonlinear inductors can be fabricated and integrated into 
classical and quantum superconducting circuits. 
By measuring transport properties such as critical current density, kinetic 
inductance, and inductive and dielectric loss, systematic studies of film 
properties as a function of particle size, packing and composition can be 
undertaken. 
Importantly, not only will this allow to optimize granular film properties 
for quantum electronics applications, but it will provide data based on accurately 
controlled variables to gain actual microscopic understanding of the origin 
and regularities behind the beneficial properties of granular superconductors.

In subsequent steps, one could proceed to the development of films based on 
size-selected clusters with protected shell structure, closing the connection 
with gas-phase studies described above and aiming to develop higher-$T_{\rm c}$
superconducting circuits as well as Josephson current-carrying 
networks~\cite{OvchinnikovPRB2012}.

\subsection*{Acknowledgement}
The author acknowledges support from the U.S. National Science Foundation under 
Grant No. DMR-2003469.

\section*{Molecular beam magnetic resonance of metal clusters with a dozen 
of atoms: Almost impossible?}
\label{Schafer}
\noindent
Jannik Mehmel and Rolf Schäfer\\

\textbf{The short-note describes the difficulties involved in conducting molecular 
beam resonance experiments on metal clusters with a dozen atoms. 
This is discussed in the context of spin relaxation due to spin-rotation coupling. 
Ideas to realize such experiments in the future are presented.}

\subsection*{Introduction}
Following the discovery of space quantization by O. Stern and W. Gerlach~\cite{stern1925}, 
Rabi's introduction of the molecular beam resonance method was decisive for the 
high-precision determination of the magnetic moments of atomic systems~\cite{rabi1939}. 
Although the methodology could also be applied to molecular systems~\cite{ramsey1950}, 
Rabi experiments on metal clusters consisting of a dozen atoms are still pending, with 
Li clusters being the only exception~\cite{hishinuma1996}. 
Such experiments would not only be sensitive tests for the cluster geometry, but also 
for the electronic structure via the determination of hyperfine interactions. 
This short note aims to explain why the Rabi experiment has failed so far and to 
outline ideas that would make such investigations possible in principle in the future.

\subsection*{Prerequisite - Refocusing}
\begin{figure*}[t]
    \centering
    \includegraphics[width=\linewidth]{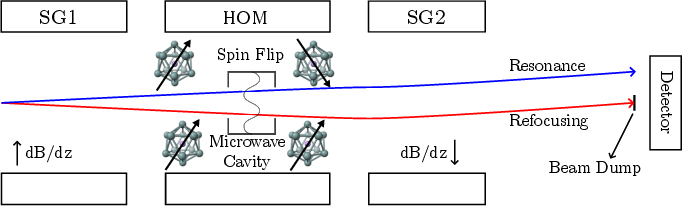}
    \caption{Scheme of the Rabi experiment with two Stern-Gerlach (SG) magnets as well 
		as a magnet with a homogeneous (HOM) magnetic field with an integrated microwave cavity. 
		The field gradient $\nicefrac{\dd B}{\dd z}$ of both SG magnets points in opposite 
		directions. 
		This leads to refocusing of a cluster if its spin state does not change (red beam 
		path). 
		Refocused clusters are blocked by a beam dump preventing the clusters to reach the 
		detector system. 
		However, if a spin transition by microwave absorption is induced the cluster 
				gets defocused and may reach the detector (blue beam path), i.e. the beam 
				dump aids to separate the two fractions of refocused and defocused clusters. 
				Measuring the intensity of the defocused fraction in dependence of the microwave 
				frequency $\nu$ or the field strength $B$ therefore allows to determine the 
				absorption spectrum. 
				It is worth noting, that clusters which spontaneously change their spin state 
				without any microwave absorption taking place will also be defocused. 
				This leads to the background signal (see main text).}
    \label{fig:rabisetup}
\end{figure*}

To conduct a magnetic resonance experiment, the spins of the clusters must be polarized 
with the aid of a first Stern-Gerlach (SG) magnet in order to then refocus the clusters 
back to their original position with the aid of a second SG magnet with a reverse gradient 
direction. 
If a change in the spin state occurs between the two deflectors due to microwave 
absorption in a homogeneous magnetic field, the particles become defocused, which can 
be observed experimentally and allows conclusions to be drawn about the microwave 
resonance (Figure~\ref{fig:rabisetup}).
However, in order for the clusters to be refocused, the spin state must not change 
without microwave absorption when passing through the two deflectors. 
But why should the spin state of isolated clusters in the molecular beam change without 
microwave absorption? 
In addition to the spin $S$, the rotation of the clusters as a whole must also be taken 
into account. 
Since the clusters can rotate freely, each rotational state with the total rotational 
quantum number $R$ splits into $2S+1$ states within the magnet. 
As a result of the magnetic deflection the $z$-component of the magnetic flux density 
$B_z$ changes slightly as the clusters move through the magnetic deflector. 
Therefore, the clusters pass through so-called avoided state crossings when traveling 
through the magnet. 
This refers to the crossing of two states that have the same value for the $z$-component 
of the total angular momentum $J_z$, i.e., the sum of the z-component of the spin 
angular momentum $S_z$ and the rotational angular momentum $R_z$~\cite{xu2008}. 
Therefore, this crossing of two quantum states is avoided and the value of $S_z$ can 
change when passing through the crossing because $R_z$ also changes at the same time, 
i.e., a spin flip can occur while maintaining the total angular momentum 
($\Delta S_z =-\Delta R_z$)~\cite{knickebein2004}. 
Whether the spin state of the clusters is preserved, depends on the number of avoided 
crossings passed through.
The number of avoided crossings passed through increases with increasing spin quantum 
number $S$ and particularly strongly with decreasing symmetry and increasing 
vibrational excitation~\cite{fuchs2019}. 
Even for rigid clusters with $S=\nicefrac{1}{2}$, more than 10 avoided crossings are 
passed through in a typical SG experiment if they possess a non-spherical symmetry. 
Whether a spin flip occurs at one of these traversed avoided crossings now depends on 
the corresponding transition probability~\cite{zener1932}. 
In addition to the velocity at which the clusters pass through the magnets, this 
depends crucially on the spin-rotation coupling. 
The most important contribution to spin-rotation coupling results from the spin-orbit 
interaction~\cite{vleck1951}. 
If spin-orbit effects are NOT negligible, the vast majority of avoided crossings are 
traversed adiabatically, i.e., with a spin flip taking place. 
In summary, this means that refocusing of clusters is only possible if, on the one 
hand, the number of avoided crossings is low and, on the other hand, the crossings are 
traversed without a spin flip. 
This is only possible if the clusters are spherical rotors, have a small spin quantum 
number, are not vibrationally excited, and if spin-orbit effects are absent. 
Therefore, single magnetically doped main group metal clusters are promising candidates 
for successful refocusing. 
This is because endohedral cluster structures with high symmetry can be 
formed~\cite{cui2007}. 
In addition, there is only one paramagnetic center, so that the total spin is low. 
And the dopant atom used can be adjusted so that spin-orbit effects are negligible. 
Therefore, it has been possible to refocus the 
\textsuperscript{55}Mn@Sn\textsubscript{12} cluster with I\textsubscript{h} symmetry 
under cryogenic conditions~\cite{fuchs2018}. 
However, it was necessary to ensure that the value of the magnetic flux density $B_z$ 
remained as constant as possible when the clusters travel through the magnets. 
If $B_z$ changes between the two deflectors, spin flips occur due to the variation 
of the magnetic field~\cite{fuchs2018}.

\subsection*{Present Challenges}
\begin{figure}[h!]
    \centering
    \includegraphics[width=0.5\linewidth]{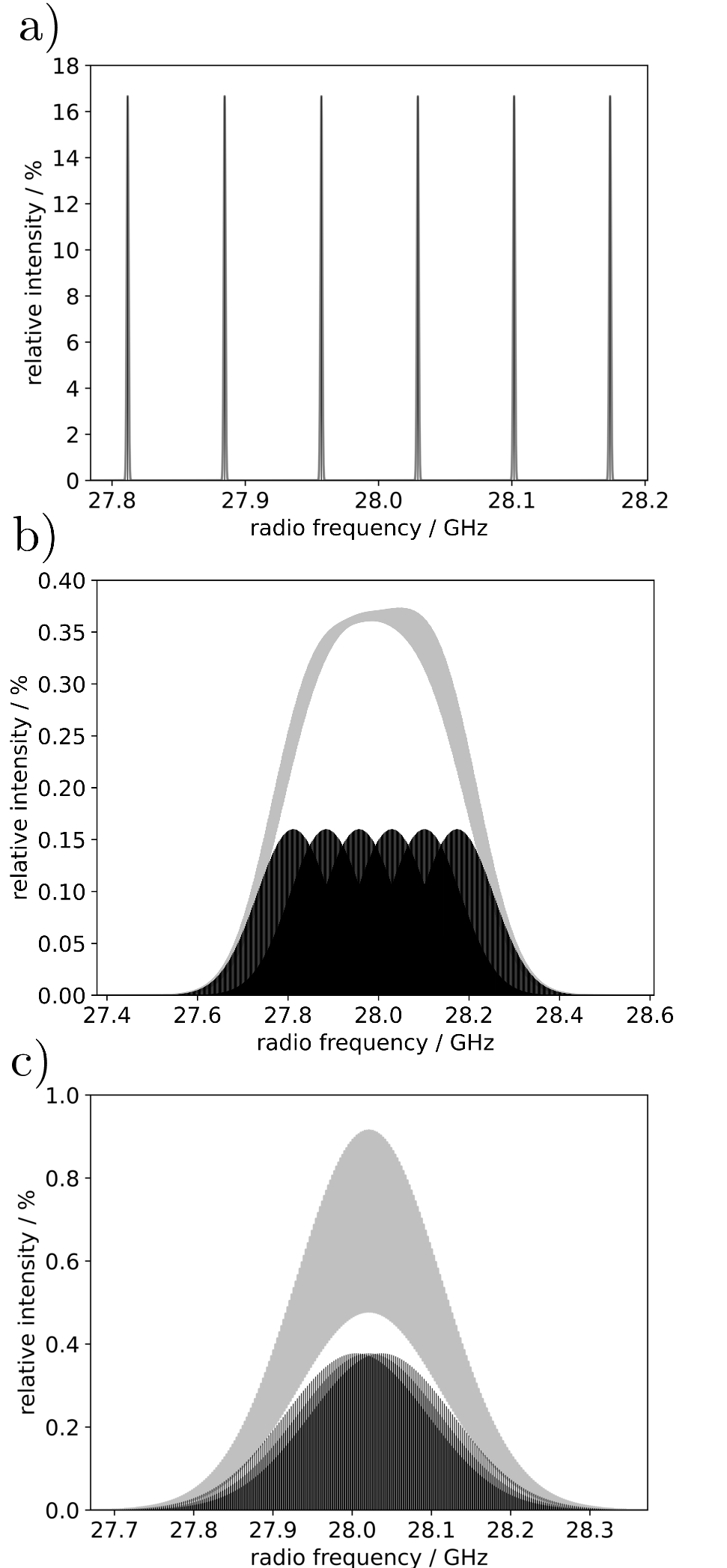}
    \caption{Simulated magnetic resonance spectrum based on equation~\ref{eq:1} 
		with $B=1~\mathrm{T}$ for a) \textsuperscript{55}Mn@Sn\textsubscript{12} 
		without spin-rotation-coupling b) \textsuperscript{55}Mn@Sn\textsubscript{12} 
		with spin-rotation-coupling and c) \textsuperscript{14}N@C\textsubscript{60} 
		with spin-rotation-coupling. 
		For details see the Appendix. 
		Black lines, which represent the stick spectrum of the individual transitions, 
		are convoluted to the gray spectrum, taking the finite width of the resonance 
		lines into account.}
    \label{fig:rabi}
\end{figure}

Although refocusing of \textsuperscript{55}Mn@Sn\textsubscript{12} has been successful, 
it has not yet been possible to detect microwave absorption. 
For this cluster, a total of six resonance lines are expected to appear as a result 
of the nuclear spin of \textsuperscript{55}Mn@Sn\textsubscript{12} (see Figure \ref{fig:rabi}a).
If all refocused clusters are prevented from reaching the detector, only when the 
resonance conditions are met should an intensity of \textsuperscript{55}Mn@Sn\textsubscript{12} 
other than zero be detectable at the detector, namely $\nicefrac{1}{6}$ of the 
total original intensity, since the cluster beam is unpolarized before entering the 
magnetic field. 
This should be measurable, even though the signal-to-noise ratio is significantly 
worse with pulsed laser evaporation sources than with continuously operated cluster 
sources. 
The fact that this has not yet been achieved is partly due to non-resonant spin flips, 
which prevent measurements from being taken against a “zero” background. 
The spin states with $S_z=\pm \nicefrac{5}{2}$ are particularly affected by this, as 
they show the greatest deflection in the deflectors. 
In addition, a small proportion of the \textsuperscript{55}Mn@Sn\textsubscript{12} 
clusters are still vibrationally excited, as the clusters are not completely thermalized 
by the carrier gas He. 
This fraction does not exhibit super-atomic behavior and also increases the background, 
thus making it difficult to detect spin resonance. 
However, the main cause of failure considering \textsuperscript{55}Mn@Sn\textsubscript{12} 
is spin-rotation coupling, not only because it leads to spin flips at avoided crossings 
and thus increases the background, but also because spin-rotation coupling causes a 
splitting of the six resonance lines. 
If the hyperfine interaction $a$(Mn) of the electron spin $S$ with the nuclear spin 
$I$(\textsuperscript{55}Mn) of manganese is taken into account up to the second order, 
assuming large magnetic fields, the energy $h\nu$ for transitions with $\Delta S_z=\pm1$ 
and $\Delta I_z=0$ as well as $\Delta R_z=0$ is given by~\cite{lips2000}
\begin{align}
\label{eq:1}
h\nu =\;  g\mu_\mathrm{B}B_z + a_\mathrm{hfi}M_I 
+ \frac{a_\mathrm{hfi}^2}{2g\mu_\mathrm{B}B}
\left[I(I+1) - M_I^2\right] + \Delta_\mathrm{SR} R_z
\end{align}
The last term captures the influence of the spin-rotation coupling $\Delta_\mathrm{SR}$. 
By analyzing the refocusing behavior, it was possible to estimate the value of 
$\Delta_\mathrm{SR}$ for \textsuperscript{55}Mn@Sn\textsubscript{12} to be 
\num{8e-9}~eV~\cite{fuchs2018}. 
The isotropic $g$-value was determined from the Stern-Gerlach experiment to be 
$2.0\pm 0.1$~\cite{hishinuma1996}. 
For typical rotational temperatures of 10~K~\cite{fuchs2021}, this leads to an 
increase in intensity below 1\%, as shown in Figure \ref{fig:rabi}b, if for the 
hyperfine interaction constant $a$(\textsuperscript{55}Mn)~=~\SI{72.42}{\mega\hertz} 
the value of the isolated Mn atom~\cite{basar2003} is taken into account and values 
of 1.5~MHz (\num{6e-9}~eV) from an atomic beam resonance experiment are used for the 
width of the resonance lines~\cite{fuchs2021_2}. 
Since, due to the very small rotational constants, even at 10~K approximately 100 
rotational states are still noticeably thermally occupied, the observable intensity 
increase for a microwave transition of a given value of $I_z$ is distributed among 
the approximately 100 different rotational states, so that only an intensity increase 
of significantly less than 1\% remains. 
And due to the existing background, this has not yet been measurable for 
\textsuperscript{55}Mn@Sn\textsubscript{12}.

\subsection*{Future Directions}
However, this analysis shows for which clusters a Rabi experiment could be successful 
in the future. 
In order to reduce the background, it would be important to thermalize the clusters 
even better under cryogenic conditions so that no vibrationally excited clusters are 
present~\cite{MiaoRSI2022}. 
Studying clusters with lower spin quantum numbers would also reduce the number of 
avoided crossings. 
However, it might also be conceivable to remove the fractions with 
$\abs{S_z}>\nicefrac{1}{2}$ from the experiment via a slit geometry. 
In addition, it would be important for the rotational constant of the clusters under 
investigation to have a value as high as possible, because this reduces the number 
of avoided crossings and fewer rotational states are thermally excited. 
It would also be very helpful to know the value of the $g$ factor as accurately as 
possible, because this would allow the range in which the microwave resonance occurs 
to be narrowed down and the signal-to-noise ratio to be improved over longer measurement 
times in this range. 
Compared to \textsuperscript{55}Mn@Sn\textsubscript{12}, endohedral-doped fullerenes 
would therefore be interesting candidates~\cite{knapp1998}. 
For \textsuperscript{14}N@C\textsubscript{60}, the rotational constant is approximately 
40\% greater than that of \textsuperscript{55}Mn@Sn\textsubscript{12}, the spin 
quantum number $S$ is only $\nicefrac{3}{2}$ instead of $\nicefrac{5}{2}$ and the 
nuclear spin quantum number $I(^{15}N)=1$ instead of $\nicefrac{5}{2}$ for 
\textsuperscript{55}Mn. 
In addition, the $g$-factor of the cluster dissolved in CS\textsubscript{2} is 
known very precisely and close to the value of the free electron~\cite{can2018}. 
This means that the measurement range is already well defined and spin-orbit effects 
are negligible. 
A simulation of the expected spectrum for \textsuperscript{14}N@C\textsubscript{60} is 
shown in Figure \ref{fig:rabi}c. 
The value of $\Delta_\mathrm{SR} = $ \num{1.1e-8}~eV for 
\textsuperscript{14}N@C\textsubscript{60} was calculated using a classical model of 
a rotating charge distribution~\cite{vleck1951}. 
This approach has yielded values for $\Delta_\mathrm{SR}$ that agree well with 
refocusing experiments for clusters that do not show significant spin-orbit 
effects~\cite{fuchs2021_3}.

It was still assumed that the rotation temperature is 10~K. 
In addition, the measured hyperfine interaction of \textsuperscript{14}N@C\textsubscript{60} 
was considered~\cite{can2018}. 
Compared to \textsuperscript{55}Mn@Sn\textsubscript{12}, the intensity is increased 
by a factor of 3, with an expected less noisy background. 
If the signal-to-noise ratio is still insufficient one could try to prepare colder 
clusters or reduce the resolution in order to detect multiple transitions with 
different $R_z$ values together. 
Also experiments with a continuously operated cluster source would be conceivable 
for endohedral-doped fullerenes~\cite{kilcoyne2010}. 
This would then allow phase-sensitive detection techniques to be used to further 
improve the signal-to-noise ratio~\cite{hishinuma1996}. 
Compared to the investigations of fullerenes in solution, this would enable the high 
resolution spectroscopic investigation of isolated clusters. 
Highly accurate numbers for the $g$ factor, precise values of the hyperfine 
interaction constant and the spin-rotation coupling would provide detailed insights 
into the geometric and electronic structure of the clusters without having to take 
into account the influence of a solvent or matrix and would also be ideal benchmarks 
for high-precision quantum chemical calculations~\cite{haase2020}.

\section*{Appendix}
To simulate the magnetic resonance spectra according to equation~\ref{eq:1} several 
parameters are needed. For \textsuperscript{55}Mn@Sn\textsubscript{12} a g-factor 
of 2.0, a hyperfine-coupling constant of \SI{72.42}{\mega\hertz}, a rotational 
constant of \SI{60.4}{\mega\hertz} and a spin-rotation coupling constant of 
\SI{8e-9}{\electronvolt} (\SI{1.9}{\mega\hertz}) is used. 
For \textsuperscript{14}N@C\textsubscript{60} a g-factor of 2.00204~\cite{can2018}, 
a hyperfine-coupling constant of \SI{15.9}{\mega\hertz}, a rotational constant 
of \SI{84.5}{\mega\hertz} and a spin-rotation coupling constant of 
\SI{1.1e-8}{\electronvolt} (\SI{2.7}{\mega\hertz}) is used. 
The spin-rotation coupling constant is computed by ~\cite{xu2008}
\begin{gather*}
    \Delta_\mathrm{SR} = \frac{\mu\mu_0e}{r^2}\sqrt{\frac{15k_\mathrm{B}T_\mathrm{rot}}{2m}}
\end{gather*}
in which $\mu$ is the magnetic moment ($\mu = g\mu_\mathrm{B}\sqrt{S(S+1)}$) of the 
cluster, $m$ and $r$ are the mass and radius of the cage. 
In order to account for spin-rotation coupling, the thermal occupation of rotational 
levels has to be considered. More specifically, the Boltzmann probability for the 
system to be in a state with a specific $R_z$ has to be computed. 
This is done by numerically computing the partition function for a spherical rotor 
for a rotational temperature of \SI{10}{K}:
\begin{gather*}
    q = \sum_{R=0}^\infty(2R+1)^2\exp(-B_\mathrm{rot}R(R+1)/(k_\mathrm{B}T))
\end{gather*}
The rotational energy levels of a spherical top are $(2R+1)^2$ degenerate. 
This degeneracy arises because the energy is independent of both quantum numbers 
$R_z$ (which describes the projection of the total angular momentum along the 
laboratory-fixed axis) and $K$ (which describes the projection along one of the 
cluster's principle axes). 
Therefore, the probability for the cluster to be in a state with a specific value 
of $R_z$ can be computed by:
\begin{gather*}
    p(R_z) = \frac{1}{q}\sum_{i = R_z}^\infty(2i+1)\exp(-B_\mathrm{rot}i(i+1)/(k_\mathrm{B}T))
\end{gather*}
These probabilities serve as weights for the different spectral transitions in 
the magnetic resonance spectrum, which can be seen as a stick spectrum in black 
in figure~\ref{fig:rabi}. 
The individual lines are convoluted with a gaussian with a FWHM of 
\SI{1.5}{\mega\hertz}~\cite{fuchs2021_2} to yield the gray overall spectrum in 
figure~\ref{fig:rabi}. 
In the Rabi experiment the clusters, which do not absorb the microwave radiation, 
are blocked. 
The ordinates in figure~\ref{fig:rabi} show the fraction of the clusters which are 
transmitted to the detector because the resonance condition is met. 

\subsection*{Acknowledgment}
This work was funded by the Deutsche Forschungsgemeinschaft (DFG, German Research 
Foundation) – Grant No. DFG-SCHA885/16-2.

\section*{Cluster Science and the Foundations of Quantum Mechanics}
\label{Pedalino}
\noindent
Sebastian Pedalino, Bruno E. Ram\'irez-Galindo, Richard Ferstl, 
Severin Sindelar, Stefan Gerlich, and Markus Arndt\\

\textbf{Recent experiments have demonstrated how cluster technologies can enable 
novel tests of the foundations of quantum physics and how they allow us to set 
bounds on any possible non-linearity of quantum mechanics. 
On the other hand, the high spatial resolution and force sensitivity of advanced 
matter-wave interference experiments are uniquely positioned to examine cluster 
properties that may otherwise be difficult to access using classical beam methods. 
We envisage a bright future for research at the interface between quantum and 
mesoscopic cluster science. We also point to universal preparation, control, and 
detection technologies that need further development for particles in the 
100~kDa – 100~MDa mass range.}

For more than a century, quantum physics has been the most accurate theory of 
nature and the basis for many modern-day technologies. 
And yet, quantum mechanics is still often seen as paradoxical. 
Its roots lie in de Broglie's hypothesis that massive bodies are described by a 
wave function, which for composite objects can extend even beyond the size of the 
particles themselves. 
It challenges our understanding of reality, locality and space-time, fueling 
ongoing debates in both physics and philosophy. 
One way to probe these aspects experimentally is to study quantum superposition 
states with particles of increasingly larger mass and complexity. 
This research program has been pushed forward at the University of Vienna, starting 
with various far-field diffraction experiments~\cite{arndt_waveparticle_1999,
juffmann_real-time_2012,brand_atomically_2015} and near-field interferometers 
for complex molecules~\cite{brezger_matter-wave_2002,
gerlich_kapitzadiractalbotlau_2007,haslinger_universal_2013,fein_quantum_2019}. 
It is currently being pursued in cluster interferometry, where we aim to observe 
quantum states of nanoparticles in the MDa range~\cite{pedalino_exploring_2022, 
kialka_roadmap_2022, pedalino_experimental_2023, pedalino_probing_2025}. 

\begin{figure*}[h]
    \centering
        \includegraphics[width=0.8\linewidth]{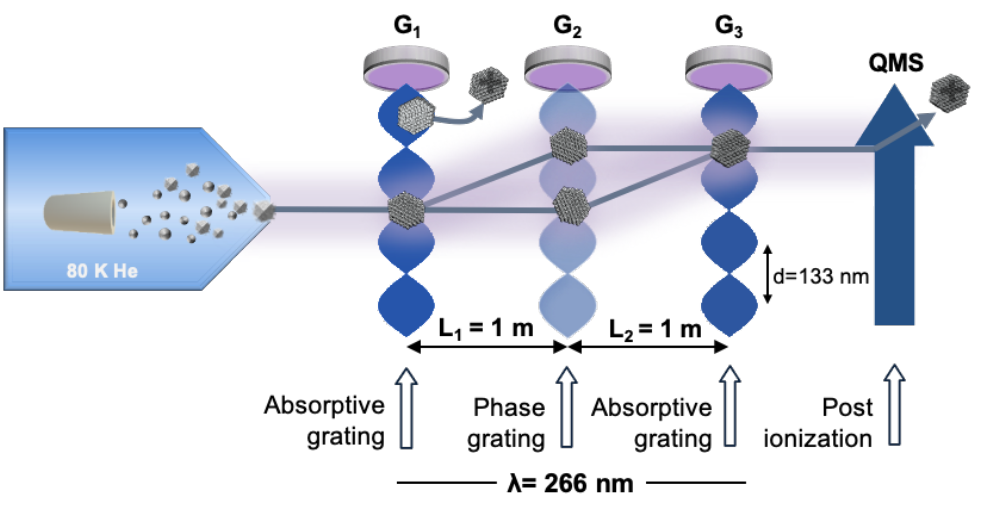}
\caption{Schematic of a Kapitza-Dirac Talbot-Lau matter-wave interferometer with 
a metal cluster source. The interferometer employs three gratings, each realized 
by standing light waves. 
The first grating is always an absorptive grating, designed to prepare the coherence 
of the matter-waves. In this configuration, depletion is achieved via photoionization. 
The second grating can act as either an absorptive or a phase grating, where the 
electric field of the standing light wave interacts with the polarizability of the 
clusters. 
The third grating functions as a transmission mask, providing spatial resolution, 
and is necessarily absorptive. 
The transmitted clusters are then ionized, mass-filtered, and counted as a function 
of the third grating's position, ultimately revealing the interference pattern.}
\label{fig:Interferometer}
\end{figure*}

Our current near-field matter-wave experiments are composed of a particle source, 
a three-grating interferometer, and a mass selective detector. 
Clusters turn out to be appealing candidates for such quantum experiments, offering 
intense neutral beams of diverse materials with varying densities and sizes, as 
well as electrical, magnetic and thermal properties. 
Most metal clusters and many dielectric nanoparticles are susceptible to efficient 
photoionization when exposed to UV laser light~\cite{reiger_exploration_2006}, 
which allows for manipulating them in a coherent way using a photodepletion 
grating~\cite{kialka_roadmap_2022}. 
For example, clusters can be ionized and extracted at the antinodes of a standing 
light wave, generating the optical equivalent of a nanoscale mechanical diffraction 
mask. 
This type of matter-wave beamsplitter finds broad applicability for clusters because 
it exploits above-threshold photo-ionization and does not require any specific 
optical line or particle resonance, a quality that is hard to find in carbon-based 
or organic molecular systems.

\subsection*{Quantum interferometry for cluster physics}\label{sec2}
Near-field matter-wave interferometry has been demonstrated to serve as an 
interesting quantum-enhanced sensor~\cite{fein_quantum-assisted_2020} for forces 
as small as $10^{-26}$~N, due to the nanoscale fringes that emerge in such 
devices~\cite{gerlich_kapitzadiractalbotlau_2007, haslinger_universal_2013} and 
that can be known with high accuracy. This opens new avenues to measure inertial 
forces and electrical or magnetic susceptibilities within electrical, magnetic, 
and optical fields, along with their gradients, to determine static and optical 
electrical or magnetic susceptibilities and absorption cross 
sections~\cite{fein_quantum-assisted_2020, tuxen_quantum_2010, gring_influence_2010, 
eibenberger_electric_2011, eibenberger_absolute_2014, mairhofer_quantum-assisted_2017, 
fein_improved_2019, yfein_quantum-assisted_2020, fein_coriolis_2020}. 
Through these properties, one can learn about cluster structure and structural 
changes~\cite{gring_influence_2010} in free flight as a function of size or internal 
temperature. With their heightened sensitivity, quantum-assisted measurements are 
emerging as a valuable complement to classical beam deflectometry, enabling access 
to particle properties that are otherwise hard to access using conventional 
techniques~\cite{antoine_direct_1999, rayane_static_1999, compagnon_molecular_2001, 
compagnon_permanent_2001, compagnon_vibration_2002, becker_dielectric_1996, 
schafer_polarizabilities_1996, schafer_electric_2008, rohrmann_stern-gerlach_2013, 
de_heer_collective_1987, selby_photoabsorption_1991, kresin_collective_1992, 
bonin_electric-dipole_1997, tikhonov_measurement_2001, 
wong_temperature-dependent_2002, moro_electric_2006, moro_electrostatic_2007, 
merthe_electrostatic_2016, tiggesbaumker_static_1996, gerlich_matter-wave_2008}.  

\subsection*{What needs to be developed?}\label{sec3}

\subsubsection*{Universal sources}
If one were to ask about our ultimate dream for a cluster quantum lab, our 
answer would be simple: a `universal cluster emitter'. 
This instrument should emit clusters of any material – metals or semiconductors, 
precisely tailored to our choice, with atomically known composition. 
While the vision of a universal cluster emitter remains an ambitious dream, the 
practical requirements for current experimental needs are more grounded yet 
still demanding. 
The mass distribution should be defined as better than $10~\%$, 
for clusters up to 100 MDa. 
A high flux is crucial, ideally more than a million neutral clusters per second, 
emitted within a solid angle of less than 1~$\mu$sr. 
The source should maintain stability over several hours to produce clear interference 
fringes. Many established cluster sources generate particles with velocities 
well above 200 m/s and log-normal size distributions with thermal or supersonic 
velocity spread. However, experiments with higher mass typically will profit 
from a lower velocity. This is true for classical deflectometry, where the 
deflection angle grows with $v^{-2}$ and also for quantum experiments, as the 
de Broglie wavelength $\lambda_\text{dB}$ scales with $v^{-1}$. 
For instance, in our current configuration, a 10~MDa cluster must travel at speeds 
as low as 2~m/s to remain compatible with existing quantum experiments. 
Equally important is tight velocity selection, with a spread better than 
$\Delta v/v <10~\%$ being highly desirable to avoid reduction of the interference 
contrast due to velocity-dependent de-phasing effects. The challenge today is 
comparable to the situation in laser physics 30 years ago, where many individual 
solutions were known before a frequency comb was invented to cover the full 
spectral range with utmost resolution in a single system~\cite{udem_absolute_1999, 
jones_carrier-envelope_2000}, or in atomic physics, where a whole field of 
research was revived when cold and quantum degenerate sources became available. 
If such a highly controlled and intense source existed also for clusters, it 
would open a plethora of fundamental experiments in physics and new opportunities 
for soft landing in catalysis and materials science, physical chemistry, and 
lithography. The challenge is substantially larger than it used to be for atoms, 
as one cannot necessarily rely on well-defined internal states, and the diversity 
of materials is virtually unbounded.

\subsubsection*{Center of mass cooling and internal state control}
Reducing the kinetic and condensation energy of particles formed in aggregation 
sources is a significant challenge. Dense cooling gas flows can reduce velocity 
spreads to below $10~\%$ but often leave particles with significant residual speeds. 
Techniques like near-effusive buffer gas sources can slow massive 
particles~\cite{patterson_slow_2015}, while immersion in superfluid helium droplets 
achieves extremely low internal temperatures~\cite{toennies_superfluid_2001}. 
As of today, achieving the combination of both slow velocities $<10$ m/s and 
ultra-cold internal temperatures of 380~mK in an intense beam of massive, neutral, 
size-selected clusters remains an intriguing challenge.  	
One may also consider combining cluster science with recent advances in 
optomechanical cooling~\cite{delic_cooling_2020, piotrowski_simultaneous_2023}, 
aiming eventually at even translational temperatures around 10~$\mu$K. 
The field of levitated optomechanics has been growing rapidly in recent years, 
so far focused on particles beyond 1 GDa. Optomechanics exploits light 
scattering, optical dipole forces or at least optical information retrieval 
about the position and velocity of nanoparticles. Since the Rayleigh scattering 
cross section scales with the square and homodyne or heterodyne detection cross 
sections scale linearly with the optical polarizability, clusters below 
10 MDa must still be considered small in this context. Consequently, they are 
hard to detect or cool by optical means alone, particularly when the goal is 
to avoid excessive internal heating.  And even if all this can be done, the 
challenge remains to scale these techniques up to beams of high brilliance.

\subsubsection*{Detection schemes}
Detecting and characterizing neutral clusters is a grand challenge at high mass 
and complexity. Upon ionization, the heavy ions suffer from broadened energy 
spreads and reduced velocities, which impair the selectivity and resolution of 
commercial mass spectrometers and lower detection efficiency due to diminishing 
secondary electron yield~\cite{brunelle_secondary_1997,zimmermann_producing_1994}. 
New frontiers can open with cryogenic impact detectors that are still being 
developed, including nano-electromechanical balances~\cite{sage_neutral_2015,
sage_single-particle_2018} or superconducting nanowire 
detectors~\cite{straus_highly_2023,straus_superconducting_2025}. 
Rayleigh scattering becomes again viable for large nanoparticles, but across 
the board, there is a clear need to advance selection and detection tools for 
masses between 100 kDa – 100 MDa.  

\subsubsection*{Precision metrology}
Cluster physics is often motivated by the desire to understand the transition 
between atomic, molecular, and bulk condensed matter physics. 
The desire to characterize the size dependence of optical, electrical, magnetic, 
thermal, or structural properties has driven many experiments for decades. 
Oftentimes, transitions in the bulk domain had already been found for atom 
numbers as small as $N = 50 - 100$. 
However, even studies from 30 years ago have already pointed to the fact that 
magic numbers can persist to $N > 5000$~\cite{gohlich_electronic_1990}. 
Specific magnetic transitions are only possible for large clusters, e.g., 
requiring that the particle size is 
comparable or greater than the London penetration depth in superconducting 
materials~\cite{xu_modification_2007}. 	

In our matter-wave experiments, the interaction of the nanoparticles with 
the gratings depends critically on their optical properties, such as absorption 
cross-section and AC polarizability. Therefore, accurate measurements of 
these properties for relevant cluster species and sizes are crucial for 
optimizing our experimental parameters and interpreting our results.
We envision a fruitful synergy between quantum optics and cluster science, 
where advancements in both fields drive mutual progress and respective 
technologies are pushed to their limits.
\backmatter

\subsubsection*{Acknowledgements}
We acknowledge financial support from a Gordon \& Betty Moore Foundation grant 
(GBMF10771, https://doi.org/10.37807/GBMF10771), and the Austrian Science Fund, FWF, 
within project 32542-N. S. Sindelar acknowledges a VDSP Master fellowship.

\section*{Developing Metal Oxide Clusters for Quantum Information Science}
\label{Sayres}
\noindent
Scott G. Sayres\\

\textbf{This perspective article highlights the rapid growth in understanding that has 
been achieved through recent experimental and theoretical results regarding the optical 
and ultrafast relaxation properties of sub-nanometer transition metal oxide clusters. 
Opportunities are suggested for utilizing their spin-polarized d-orbitals and related 
magnetic properties for applications in the emerging fields of molecular spintronics 
and quantum information science.}

Cluster science has provided the scientific community with model systems for fundamental 
research and foundational understandings about how defect sites affect chemical 
reactions and how bulk materials operate. 
The continuous efforts of many research groups are aimed at measuring the properties 
of clusters as they change atom by atom in the gas phase. 
However, the modern frontiers of cluster research go far beyond simply digesting 
and leveraging the vast knowledge recorded in the past few decades. 
The cluster community is also now aimed toward uncovering new phenomena that 
emerges at the sub-nanometer scale and synthesizing atomically precise materials 
and/or coatings that capitalize on their unique properties. 

Novel quantum phenomena discovered in transition metal oxides over the past few 
decades has invigorated new appreciation for employing clusters to understand the 
factors responsible for their properties. 
The electronic structure of such strongly correlated materials arises from 
competing interactions among charge, spin, orbital and lattice degrees of freedom. 
This enables some of the most intriguing yet poorly understood phenomena in 
condensed matter physics, including insulator-to-metal transitions, polaron 
formation and propagation, topological behavior, superconductivity, and 
half-metallicity. 
Such behaviors in strongly correlated materials are thought to follow a complex 
route of intermediate states, accompanied by changes in spin and lattice parameters. 
The electronic and lattice degrees of freedom can be transiently decoupled, 
making ultrafast spectroscopy a powerful technique for revealing their operation. 

Notable progress has been made in understanding the unique properties of strongly 
correlated materials through recent characterization of their electronic properties 
via optical spectroscopy methods. 
Over the past few years, my research group has made extensive progress in unraveling 
the ultrafast dynamics in a wide range of gas-phase metal oxide clusters to obtain 
molecular-level insight into bulk processes. 
Our recent studies on the excited state dynamics of neutral clusters have brought 
significant advancement in understanding the transport, localization, separation, 
relaxation, and recombination of free charge carriers, polarons, and excitons as 
influenced by size, morphology, and 
composition~\cite{D0CP03889J,doi:10.1021/acs.jpclett.1c00840,doi:10.1021/jacs.1c07275,
D4CP01544D,doi:10.1021/acs.jpca.4c05013,doi:10.1021/acs.jpclett.5c01379}.
Other groups have demonstrated the activation of specific vibrational modes through 
photoexcitation~\cite{D3CP06229E,D3CP02055J}. 
New understandings of the role of electron correlation, onset 
of metallic behavior, and angular momentum on their relaxation mechanisms can now be 
applied to the design of nano-structured materials. 
Such phenomena affect performance 
details, which impact their development as tunable materials for photocatalysis, 
spintronics, and other industrial applications. 
Management of both charge and energy transport at the molecular level is essential 
for advancing existing technologies and for the rational design of next-generation 
photoactive materials. 
Our experiments have provided the benchmarking datasets needed to expand and calibrate 
theoretical approaches that provide even deeper insight into the competition between 
excited-state relaxation and energy transport in metal oxide clusters. 
We have shown that ultrafast pump-probe spectroscopy can be utilized to disentangle 
the distinct dynamical regimes of femtosecond relaxation through the manifold of 
excited states from the significantly slower charge carrier recombination that 
operates on the picosecond timescale. 
Ab Initio nonadiabatic molecular dynamics simulations (NA-MD) have recently become 
possible for metal oxide clusters to provide unprecedented mechanistic insights 
into how excited states evolve. 
The tight synergy between experimental and theoretical data shows the relationship 
between energy gaps, nonadiabatic couplings, and densities of states that drives 
competition between excited state population spreading and coherent relaxation of 
photoexcited states of titania clusters~\cite{Recio-PooJACS2025}. 
Nevertheless, the significant variation in the relative strength of these factors, 
arising through the interplay of structure, atomic composition, and size at the 
molecular level, ensures that optical properties and ultrafast dynamics of clusters 
remain highly active directions of research. 
Advancements in laser technology now enable exploration of long-range coupling and 
charge migration in their native timescales with resolution down to the attosecond 
timescale. 
Thus, there remains considerable interest in continuing spectroscopy 
experiments to explore the rich distribution of accessible excited states within 
each cluster, each with promise for unlocking novel properties and providing new 
understandings of energy transduction. 

There is compelling evidence that the magnitude of both the spin and magnetic 
moments 
can be selected by changing the morphology, atomic composition and size of the cluster. 
The partially filled, frontier orbitals in open-shell metal oxide clusters are 
strongly spin polarized due to the ferromagnetic (FM) coupling that serves as the 
foundation for many of their interesting behaviors. 
Novel properties arising from the spin-polarized d-electrons in specific transition 
metal oxide clusters can be recruited to open entirely new research frontiers 
in spintronics. 
Promising candidates include copper oxides, which develop unprecedentedly 
large spin magnetic moments when shrunk down to the molecular 
level~\cite{doi:10.1021/acs.jpclett.5c01379}. 
We found that the local structural features, such as the formation of $\mu 2$-O sites, 
affect the dynamics and drive the high spin multiplicities and related magnetic 
phenomena within these clusters. 
Every atom has a straightforward impact on the cluster’s spin multiplicity and 
related magnetic moment, and enables linear tuning of their behaviors with 
composition~\cite{doi:10.1021/acs.jpclett.5c01379}. 
Other transition metal oxide clusters have also attracted significant attention 
with the discovery of a class of cube-like structures, termed 
“metalloxocubes”~\cite{PhysRevB.62.8500,doi:10.1021/acs.jpclett.1c04098,
doi:10.1021/acs.jpclett.3c03637}.
Such hollow structures possess electronic delocalization and cubic aromaticity, 
which are associated with enhanced stability. 

Gas phase research consistently demonstrates the importance of atomic precision 
over nearly every measurable property. 
However, in recent years, a transformational shift has occurred from thinking 
about clusters as atomically precise materials inside a mass spectrometer toward 
developing them as highly tunable components for engineering specialized coatings 
and materials. 
There is considerable interest in developing sub-nanoscale clusters as a form 
of defect engineering for tailoring the electronic, magnetic, optical, 
mechanical, and quantum properties of strongly correlated metal oxides. 
Recent innovations in synthetic approaches have enabled the scale-up of atomically 
precise nanostructures (polyoxometalates~\cite{https://doi.org/10.1002/anie.200902483,D0NA00877J} 
and noble 
metal nanoclusters~\cite{doi:10.1021/acscentsci.5c00139,10.1063/1.5090508}) 
to make meaningful impact at industrially relevant scales. 
Advancements in soft-landing deposition 
technology~\cite{doi:10.1021/nn2039565,doi:10.1021/acs.jpclett.2c03114,
https://doi.org/10.1002/cctc.201901795} provide additional 
manufacturing approaches with possibly even greater tunability and novelty. 
Thus, the variety of these synthetic techniques provides several options for 
bringing clusters into the design and verification of tangible 
materials~\cite{doi:10.1021/nn800820e}. 

Transition metal oxide clusters offer both high stabilities and extraordinary 
\mbox{spin/magnetic} properties~\cite{doi:10.1021/acs.jpclett.3c03637} that can be 
developed into faster, smaller, and more energy-efficient electronic devices. 
The tight relationship between the local magnetic moments and variation in 
structural features unique to clusters provides exquisite tunability toward 
generating specific spin states and magnetic moments for the formation of 
molecular-sized magnets. 
By understanding and controlling the magnetic and spin dependent properties, 
open-shell clusters can serve as a platform for developing molecular-scale 
spintronics and quantum information science (QIS) in computing and sensing. 
Unlike classical electronics that rely on the charge of electrons, spintronics 
is a technology that manipulates the spin state of electrons to encode and 
process information. 
The quantum mechanical properties of spin enable registers of qubits, or 
quantum bits, to reside in conventional states of $|0\rangle$ or $|1\rangle$, but can also 
become a superposition of the two. 
This new type of quantum logic enables increased information storage and 
possibly outpace calculations of classical bits comprising conventional 
computers. 
Additionally, spin-based qubits can resist decoherence, the primary process 
that degrades quantum information.
Furthermore, the array of accessible low-energy spin states within open-shell 
clusters provides the intrinsic multi-level framework needed to design qudit 
(or multiple-state qubit) architecture. 
Transitioning to qudit logic offers several advantages over binary encoding 
and therefore may greatly increases the efficiency of quantum computing. 
The larger state space to store and process information reduces the number 
of distinct physical units or overall gate count required for operations and 
therefore simplification of experimental setups. 
A single multilevel quantum object can also host robust error-corrected logical 
units within itself to combat errors and harmful noise. 
Encoding such quantum error correction (QEC) is essential to protect against 
degradation of quantum information through decoherence as the system relaxes 
to an impure state over time. 

In conclusion, this perspective article outlines a thriving direction for 
the future of cluster research for years to come by driving innovation in 
the emerging developments of spintronics, and quantum information science. 
Clusters, particularly metal oxides, are promising candidates to serve as 
qudit hosts that overcome limitations of conventional electronics and empower 
quantum computing to achieve calculations currently impossible by classical 
computers. 
This promising direction in cluster science offers exciting opportunities, 
but also appreciable challenges that must be overcome in translating such 
properties into tangible material and coating engineering. 
Before clusters can be fully utilized as building blocks for the bottom-up 
assembly of new materials that capitalize on their unique and tunable chemical 
and physical properties, more information related to the cluster-surface and 
cluster-cluster interactions is needed. Their magnetic interactions and 
spin-dependent properties also remain highly underexplored, leaving key 
knowledge gaps to address with ongoing research. 

\subsection*{Acknowledgement} 
Acknowledgement is made to the donors of the American Chemical Society Petroleum 
Research Fund for support of this work.

\section*{Boron Nanoclusters}
\label{Wang}
\noindent
Lai-Sheng Wang\\

\textbf{Size-elected boron clusters have been investigated by anion photoelectron 
spectroscopy in combination with theoretical calculations. 
These clusters are found predominantly to be planar, laying the foundation for 
2D boron (borophene). 
The issue of structural changes as a function of cluster size and the challenges of 
studying large boron clusters are discussed.}

The study of carbon clusters led to the discoveries of fullerenes and carbon 
nanotubes~\cite{KrotoNature1985,IijimaNature1991} that helped lay the foundation of 
nanoscience and the synthesis of graphene~\cite{NovoselovScience2004}. 
Are there other elements that can form similar nanostructures? Carbon’s lighter 
neighbor, boron, is a promising candidate because of the strong boron-boron bond, 
as reflected by the high melting point of bulk boron, second only to diamond or 
graphite among the main group elements, as well as the fact that many allotropes 
of boron are superhard materials~\cite{OganovJSM2009}.

\begin{figure}[h!]
    \centering
    \includegraphics[width=\linewidth]{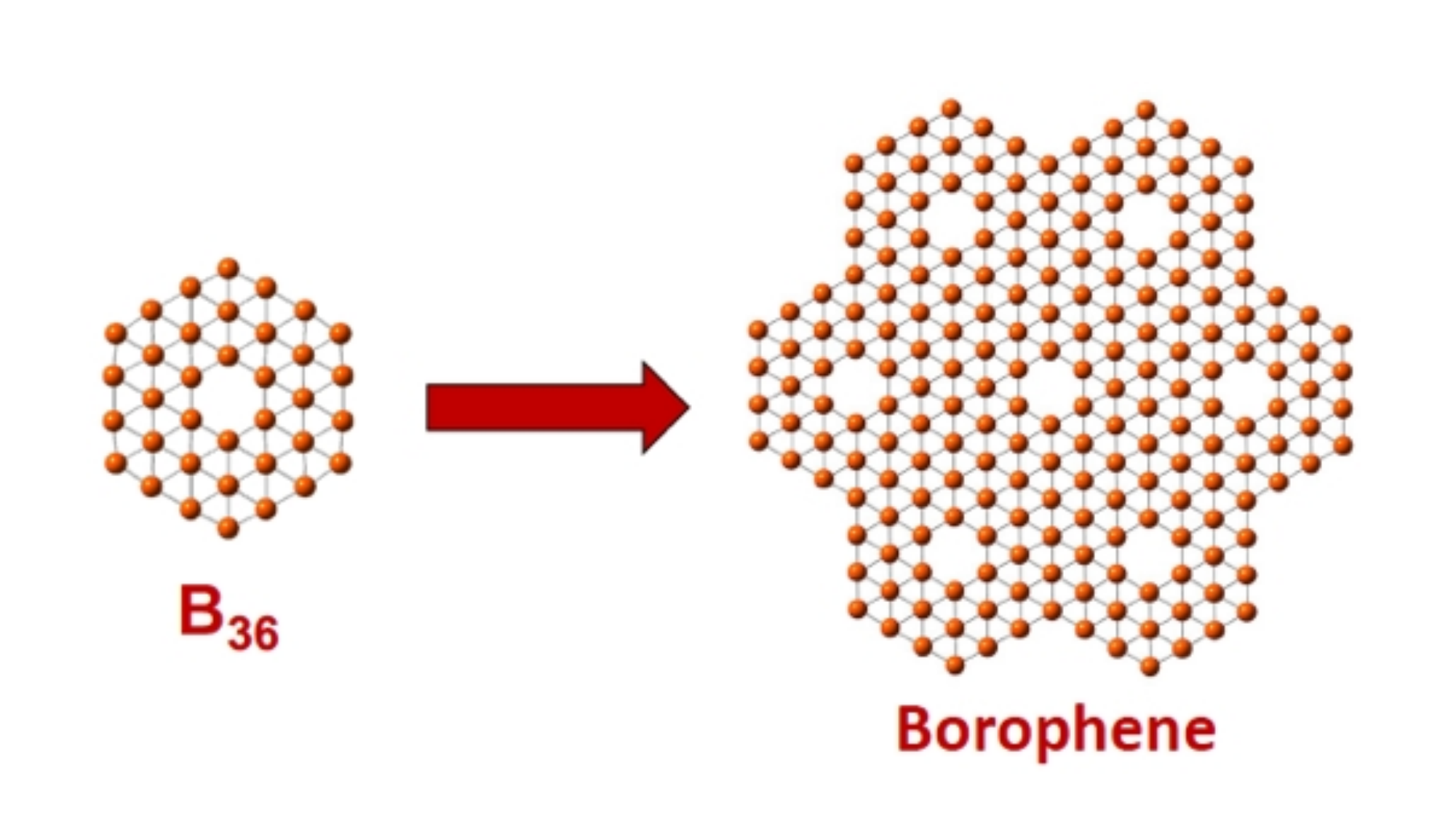}
    \caption{From planar boron clusters to borophenes.}
    \label{Wang1}
\end{figure}

Motivated by this question, we have been systematically investigating boron clusters 
as a function of size using photoelectron spectroscopy in combination with 
computational chemistry~\cite{WangIRPC2016,JianCSR2019}. 
Bare boron clusters are found to possess planar structures, in contrast to that of bulk
boron, which is dominated by three-dimensional polyhedral building 
blocks~\cite{AlbertACIE2009}. 
The propensity for planarity has been found to be a result of both $\sigma$ and $\pi$ 
electron delocalization over the molecular plane~\cite{SergeevaACR2014}, as a result 
of boron’s electron deficiency. 
The planar boron clusters laid the foundation for 2D boron. 
In particular, the B$_{36}$ cluster, which was found to have a highly stable planar 
structure with a central hexagonal vacancy (Fig.~\ref{Wang1}), provided the first 
experimental evidence that single-atom boron-sheets with hexagonal vacancies
(named borophene) are viable~\cite{PiazzaNC2014}. 
Borophenes have since been synthesized and characterized on inert
substrates, forming a new class of synthetic 2D materials~\cite{KanetiCR2022}. 
The B$_{40}$ cluster has been found to have a cage structure with $D_{2d}$ symmetry, 
the first all-boron fullerene~\cite{ZhaiNanoscale2022}. 
The largest boron cluster characterized experimentally to date, 
B$_{48}^–$, possesses a bilayer structure~\cite{ChenNanoscale2021}, suggesting the 
feasibility of bilayer borophenes~\cite{ChenNatChem2022,LiuNatMat2022}.

Despite these tremendous progresses, there are still many interesting questions 
for larger boron nanoclusters. 
Will they continue to form cage structures? 
When will the 2D to 3D transition occur? 
When will the bulk-like icosahedral B$_{12}$ building block appear? 
Ultimately, will there be stable boron nanoclusters that can be synthesized like the 
fullerenes? 
To answer these questions requires systematic investigations of the structures and 
bonding of large boron clusters as a function of size. Well-resolved photoelectron 
spectroscopic features are necessary as electronic fingerprints to compare with 
computational results. 
However, large boron clusters pose significant challenges both experimentally and 
computationally.

There are two potential issues to study large boron clusters using photoelectron 
spectroscopy or other spectroscopic techniques. 
First, because of the huge heat of formation, nascent boron clusters produced from 
a laser vaporization supersonic cluster source can be very hot and cluster
cooling becomes difficult. 
This challenge can in principle be solved by using cryogenic ion trap techniques. 
We have built cryogenically-cooled Paul traps to cool anions from an electrospray
ionization source~\cite{WangRSI2008} and a laser vaporization cluster 
source~\cite{KocherilJCP2002}. 
The latter is much more difficult because of the low number density of cluster anions 
that can be produced from a laser vaporization cluster source. 
Thus far, we have worked with smaller boron clusters from the laser 
vaporization cluster source~\cite{GaoJPCL2025}. 
With further improvement of the design, it is potentially viable to trap and cool 
large boron clusters. 
There are other linear ion traps that may also work well for large boron 
clusters~\cite{GerlichACP1992,HockJCP2012}.

The second challenge to study large boron clusters using photoelectron spectroscopy is the
potential co-existence of low-lying isomers. 
While creating cold clusters will help minimize
population of low-lying isomers, co-existence of isomers with close energies may not be
avoidable for large boron clusters. 
The co-existence of isomers will give rise to overlapping
spectral features, resulting in broad photoelectron spectra that would be difficult 
to interpret and compare with theoretical calculations. 
This problem can potentially be resolved by isomer separation techniques before 
performing the photoelectron spectroscopy experiment. 
Ion mobility has been used to separate isomers of carbon cluster anions and biological 
molecular ions for different spectroscopic 
investigations~\cite{FromherzPRL2022,VonderachAC2011,MarltonCC2022}. 
When combined with cryogenic ion traps to create cold 
ions~\cite{KamrathACR2018,BuntineRSI2022}, the ion mobility 
technique can be used to allow isomer-specific photoelectron spectroscopy to be 
conducted on large boron clusters. 
This would be an exciting development, which will not only allow large boron 
clusters to be studied, but will also open the door to study other complex cluster systems.

\subsection*{Acknowledgement}
This work was supported by the National Science Foundation (CHE-2403841).

\section*{The authors declare no competing financial interests.}


\end{document}